\documentclass[aps,pra,twocolumn,superscriptaddress,longbibliography,floatfix]{revtex4-2}
\usepackage[T1]{fontenc}
\usepackage{graphicx}
\usepackage{amsfonts}
\usepackage{bm}
\usepackage{verbatim}
\usepackage{color}
\usepackage{array}
\usepackage{ifthen}
\usepackage{braket}
\usepackage{cancel}
\usepackage{natbib}
\usepackage[colorlinks=true, allcolors=navyblue, breaklinks=true]{hyperref}
\usepackage{CJK}
\usepackage{dcolumn}
\usepackage[mathscr]{eucal}
\usepackage[normalem]{ulem}
\usepackage{subfigure}
\usepackage{epsfig}
\usepackage{xspace}
\usepackage{color}
\usepackage{bbold}
\usepackage{tabu}
\usepackage{mathtools}

\newcommand{\bea}{\begin{eqnarray}}
\newcommand{\eea}{\end{eqnarray}}

\usepackage{xcolor}

\definecolor{mpurple}{RGB}{122,93,199}
\definecolor{maroon}{RGB}{128,0,32}
\definecolor{darkred}{RGB}{139,0,0}
\definecolor{burntorange}{RGB}{191,87,0}
\definecolor{forestgreen}{RGB}{34,102,51}
\definecolor{darkteal}{RGB}{0,90,90}
\definecolor{navyblue}{RGB}{30,30,150}
\definecolor{cobalt}{RGB}{15,30,205}
\definecolor{crimson}{RGB}{170,10,60}
\definecolor{chocolate}{RGB}{105,60,20}

\newcommand{\monalisa}[1]{\textcolor{cobalt}{#1}}

\begin{document}

\title{Entanglement diagnostics of stochastic-resetting steady states: theory and measurement on superconducting qubits}

\author{Jakob Murauer}
\email{jakob.murauer@unibw.de }
\affiliation{Research Institute CODE, University of the Bundeswehr Munich, D-81739 Munich,
Germany}

\author{Sabine Tornow}
\email{sabine.tornow@unibw.de}
\affiliation{Research Institute CODE, University of the Bundeswehr Munich, D-81739 Munich,
Germany}

\author{Manas Kulkarni}
\email{manas.kulkarni@icts.res.in}
\affiliation{International Centre for Theoretical Sciences, Tata Institute of Fundamental Research,
Bengaluru 560089, India}

\author{Eli Barkai}
\affiliation{Department of Physics, Institute of Nanotechnology and Advanced Materials, Bar-Ilan University, Ramat-Gan 52900, Israel}

\author{Satya N. Majumdar}
\email{satyanarayan.majumdar@cnrs.fr }
\affiliation{LPTMS, CNRS, Univ. Paris-Sud, Université Paris-Saclay,
91405 Orsay, France}

\author{Monalisa Singh Roy}
\email{monalisasroy@gmail.com}
\affiliation{Department of Physics, Bar-Ilan University, Ramat Gan 52900, Israel}
\affiliation{PlanQC GmbH, Garching near Munich 85748, Germany}

\date{\today}

\begin{abstract}
Stochastic resetting drives quantum systems into nonequilibrium stationary states unreachable under uninterrupted unitary evolution, and their entanglement depends on whether the reset history is resolved or averaged over. We study this interplay in the transverse-field Ising chain, deriving closed-form results for two qubits and evaluating chains of up to 12 qubits exactly. In the two-qubit system, these closed-form results show that phase matching between energy eigenstates produces resonances, and their visibility in the stationary state is determined by the symmetry of the reset state. Measurements on an IBM superconducting processor show the entropies evaluated before and after the reset average closely following noiseless predictions with no fitted parameter, while concurrence lies below prediction across most of the hardware measurements. The two diagnostics respond differently because they read different components of the same reconstructed state, and a depolarizing model captures the concurrence suppression with one fitted parameter. The difference between the two entropies, i. e., the Jensen gap of the von Neumann entanglement entropy, equals the Holevo information that a subsystem retains about the time since the last reset. With a Bell-state reset, this information grows with chain length across a half-chain cut over the sizes studied, without visible saturation at the smaller reset probabilities. Since reset times are sampled and recorded classically, the trajectory-averaged entropy, and with it the Jensen gap, is obtained without postselection, in contrast to monitored circuits, where trajectory-resolved entanglement requires sampling exponentially rare measurement records.
\end{abstract}

\maketitle

\section{Introduction}
\label{sec:introduction}
{Quantum processors can reset all their qubits together to a fixed state in the middle of a circuit~\cite{2021_Corcoles_dynamic}. Such resets are a form of engineered dissipation, which can prepare a chosen quantum state~\cite{2009_Verstraete_dissipation}. Such resets applied at random times, with a fixed probability per step, are known as stochastic resetting~\cite{2011_Evans_diffusion,2020_Evans_resetting}. Every step then applies the same rule, a reset with probability $p$ and the circuit unitary step otherwise. For any $p > 0$ this drives the qubits to a unique stationary state, a nonequilibrium steady state in which the reset breaks detailed balance and sustains probability currents, independent of the state they started in~\cite{2021_Perfetto_resetting,2025_Wald_discrete}. On a processor, this state is reconstructed from many runs of the circuit, each one realization of the reset dynamics. The entanglement of the stationary mixture must therefore be distinguished from the entanglement present in the individual realizations. This distinction matters when interpreting entropy measurements, because averaging over reset histories mixes reduced states of different ages, and this mixing adds entropy that does not come from entanglement, as Sec.~\ref{sec:diagnostics} makes precise. The question then becomes how much entanglement the stationary state retains, what information can be gathered from it about the history of the reset, and how hardware noise changes the entanglement diagnostics measured on a quantum processor.}
 
Stationary states under resetting have been studied extensively for classical stochastic processes~\cite{2020_Evans_resetting}. A diffusing particle returned to its starting position at a constant rate, for example, reaches a nonequilibrium stationary state~\cite{2011_Evans_diffusion}. Stochastic resetting of quantum systems was introduced independently by Mukherjee~\textit{et al.}~\cite{2018_Mukherjee_reset} and by Rose~\textit{et al.}~\cite{2018_Rose_spectral}; the renewal equation for the density matrix of a general quantum system under resetting was derived in the former. When the underlying dynamics is unitary, the interplay between evolution and resetting can produce a nonequilibrium stationary state. Such states can retain coherences and entanglement, which depend on the Hamiltonian, the reset state, and the reset statistics. Perfetto~\textit{et al.}~\cite{2021_Perfetto_resetting} established a framework for constructing these states in closed quantum many-body systems, and Magoni~\textit{et al.}~\cite{2022_Magoni_correlations} analyzed the correlations that emerge in noninteracting systems under resetting. Both these treatments are primarily continuous in time. Realizing the same construction on a quantum processor, where its coherence and entanglement can be characterized experimentally, requires implementing resets on discrete time steps instead.
 
A generalization from continuous to discrete time was discussed by Mukherjee~\textit{et al.}~\cite{2018_Mukherjee_reset} in the context of periodically driven integrable models that are reset after a random integer number of drive cycles. More recently, Wald~\textit{et al.}~\cite{2025_Wald_discrete} developed a framework with a fixed reset probability per step, which we use here and which is convenient for gate-based devices, where each unitary step is implemented by a fixed gate sequence. At each step, the system is either returned to a fixed reset state $\ket{\psi_0}$ with probability $p$ or evolves under the fixed unitary $U=e^{-iHT}$ with probability $1-p$, where $T$ is the duration of a unitary step. The stationary state retains off-diagonal coherences at spectral resonances, where two eigenvalues of the step unitary coincide. For eigenstates $\ket{E_i}$ and $\ket{E_j}$ of $H$, with eigenenergies $E_i$ and $E_j$,
this is the condition
$ e^{-iE_iT} = e^{-iE_jT}$, or equivalently, with $\mathbb{Z}$ the set of integers,
\begin{equation}
(E_i - E_j)\,T \in 2\pi\mathbb{Z}.
\label{eq:resonance_intro}
\end{equation}
Under these conditions, the coherence between the two eigenstates survives the averaging over reset histories in full, provided the reset state overlaps with both of them, $\braket{E_i|\psi_0}\braket{\psi_0|E_j} \neq 0$ (Sec.~\ref{sec:resonances}).

The same condition also appears in the first-passage problem of quantum systems monitored by repeated measurements at a fixed time interval $T$. If the starting site of a quantum walk is measured once per interval until the walker is first found there, the mean number of measurements is an integer \cite{2013_Grunbaum_recurrence,2017_Friedman_walks}, as first observed on a quantum computer for a single qubit~\cite{2023_Tornow_walks}. This integer drops abruptly at values of $T$ for which Eq.~(\ref{eq:resonance_intro}) holds for two eigenstates that overlap the starting site \cite{2025_Yin_resonances}. Restarting the walk broadens these drops by an amount set by a time-energy uncertainty relation \cite{2025_Yin_uncertainty}. Related degeneracies produce fractionally quantized recurrence times in interacting spin systems, which has also been demonstrated experimentally on a superconducting processor \cite{2026_Liu_recurrence}. We show that the entropies of the reset stationary state have resonances at the same condition. As the exchange coupling strength or the transverse field of the Hamiltonian is varied at fixed step duration, both entropy diagnostics show narrow peaks and dips along the loci where Eq.~(\ref{eq:resonance_intro}) holds (Sec.~\ref{subsec:heatmap_diagnostics}). An entropy, a quantity not usually associated with periodic dynamics, therefore marks where two eigenvalues of the step unitary coincide, as the mean return time does.

In classical stochastic processes, resetting is known to speed up search~\cite{2011_Evans_diffusion,2020_Evans_resetting}. For example, the mean first-passage time of a diffusing particle to a target, which diverges for free diffusion, becomes finite under resetting to the initial position and passes through a minimum as the resetting rate is varied \cite{2011_Evans_diffusion}. This observation has been developed into a general theory of first passage under restart \cite{2017_Pal_restart,2020_Evans_resetting}. Restarting also reduces the mean detection time in monitored quantum walks \cite{2023_Yin_restart}, and related restart effects are found in a quantum walk with flat-band localization \cite{2026_Das_lackadaisical}. In a monitored quantum search on a complete graph, resetting the walker whenever the detector has not fired within a set interval can improve the scaling of the search time with the database size beyond that of Grover's algorithm, although the total search time, with the cost of the measurements counted, stays within Grover's bound~\cite{2026_King_search}.

Beyond first passage, Kulkarni and Majumdar showed that stochastic resetting can generate entanglement in interacting quantum systems \cite{2023_Kulkarni_entanglement}, and Ghosh~\textit{et al.}~\cite{2026_Ghosh_concurrence} found that periodically driven spin chains under resetting sustain a finite stationary pairwise concurrence above a critical reset rate. Further studies have broadened the analysis to nonanalytic stationary behavior in noninteracting many-body dynamics \cite{2025_Soldner_nonanalyticities}, emergent correlations in bosonic systems \cite{2025_Kulkarni_bosons}, local signatures of quantum many-body scar towers \cite{2026_Gotta_scars}, control of defect production across the transverse-field Ising critical point \cite{2026_Jafari_defects}, revivals of quantum correlations between qubits coupled to a reset-driven critical environment \cite{2026_Jafari_resetenv}, reset-induced quantum non-Markovianity~\cite{2026_Carollo_nonMarkovianity}, the preparation of a target ground state by periodic resetting of an ancilla coupled to the system~\cite{2024_AlcaldePuente_ancilla}, and entanglement transitions of non-Hermitian quasiparticles in monitored systems \cite{2022_Turkeshi_resetting}. Each of these studies asks whether resetting or monitoring produces a specific signature in a particular system.

A related experiment by Murauer~\textit{et al.}~\cite{2026_Murauer_NESS} realizes reset stationary states of the Floquet transverse-field Ising model for up to seven qubits and tracks their magnetization across the crossover between the ferromagnetic and the paramagnetic regime. Here we investigate the entanglement of a reset stationary state, compare what different entropy and entanglement measures reveal about it, and test how the measures respond to hardware noise.

Using analytical calculations, numerical simulations, and remote experiments on an IBM superconducting processor, we compare the averaged-state entropy $S(\bar\rho_A)$, evaluated after the reduced density matrix is averaged over reset histories [Eq.~(\ref{eq:S_rhobar})], the trajectory-averaged entropy $\bar S_A$, evaluated before this average is taken [Eq.~(\ref{eq:Sbar})], and the Wootters concurrence $C(\bar\rho_\infty)$ of the two-qubit stationary state \cite{1997_Hill_concurrence,1998_Wootters_concurrence,2009_Horodecki_review}. These measures characterize different properties. The concurrence measures entanglement in the stationary mixture, while $\bar S_A$ is the mean entanglement of the pure realizations in the ideal reset protocol. The averaged-state entropy $S(\bar\rho_A)$ measures the mixedness of the reduced stationary state and also includes the effect of averaging over reset ages. Its excess over $\bar S_A$, which is called the the Jensen gap of the von Neumann entanglement entropy, is the extra uncertainty about subsystem $A$ that comes only from not knowing the time since the last reset. It equals the Holevo information [Eq.~(\ref{eq:holevo})], which bounds how much a measurement on $A$ alone can reveal about that time.

The distinction between entropy before and after ensemble averaging is central to measurement-induced dynamics \cite{2019_Skinner_MIPT,2019_Li_hybrid,2020_Gullans_purification,2026_Hamazaki_review,2020_Turkeshi_criticality,2021_Turkeshi_clicks}. A related transition, driven by reset channels alone rather than by measurement, has also been found in random quantum circuits \cite{2026_Yokoyama_resetcircuits}. Direct reconstruction of individual trajectories in monitored circuits can require postselection on complete measurement records, motivating the development of postselection-free probes \cite{2025_Kamakari_XEB}. In the ideal reset protocol, each realization is specified by its age, i.e., the number of unitary steps since the most recent reset, which is just a classical variable. States of a given age can therefore be reconstructed from repeated measurements, and their entropies averaged with the known age distribution. This avoids postselection on complete measurement records, although resolving the different ages requires additional measurement statistics.

We study the transverse-field Ising model under this discrete-time protocol, combining an exact treatment of two interacting qubits with numerical calculations for chains of up to $N=12$ sites. For two qubits, we implement the protocol on an IBM superconducting processor (Secs.~\ref{sec:model} and \ref{sec:diagnostics}), obtain the two entropies and concurrence from quantum state tomography, and compare them with exact calculations based on the renewal series. The two-qubit spectrum gives four families of phase-matching loci, while reset-state overlaps and parity determine which eigenstate pairs contribute to the averaged stationary state (Sec.~\ref{sec:reduced_density_matrix}). The trajectory-averaged entropy shows further features, for instance where the realizations alternate between two pure states, that leave no resonance in the averaged state. On hardware, both entropies closely follow the noiseless theory with no fitted parameter  (Sec.~\ref{sec:diagnostics}), whereas concurrence lies below it across most of the experimentally studied range of parameters. A depolarizing model with a single fitted strength captures the overall concurrence suppression at $p=0.05$; a first-order calculation compares the noise responses of averaged-state entropy and concurrence (Sec.~\ref{sec:noise}). Their different responses show why agreement of the reduced-state entropy with noiseless theory does not by itself establish that the stationary-state entanglement is equally well preserved.

The chain calculations in Sec.~\ref{sec:exact_chain} examine the Jensen gap of the von Neumann entropy for single-site and half-chain subsystems. The single-site gap is bounded by one bit; for $p$ from 0.05 to 0.65 at $\alpha = 0.2$ and 2.0, the half-chain gap grows for the tiled-Bell reset, without visible saturation at the smaller reset probabilities up to $N=12$. These finite-size results show how the information about reset age depends on the subsystem and reset state. The gap is also capped by the Shannon entropy of the age distribution, which is finite at any nonzero reset probability and does not grow with $N$. The chain extension therefore examines how the distinction between the two entropies develops beyond two qubits, while the hardware measurements establish experimental access to the two-qubit quantities.

\begin{figure}
   \begin{center}
\includegraphics[width=1.0\linewidth]{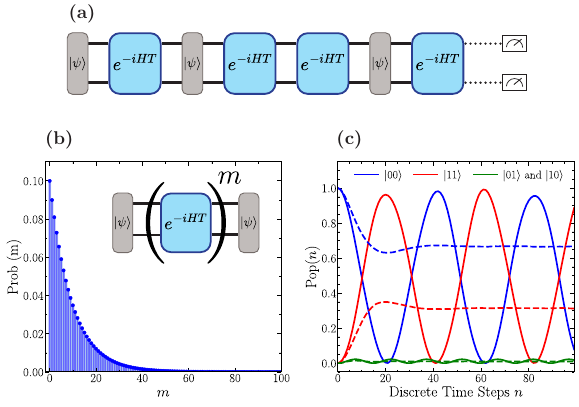}
    \end{center}
\caption{(a) Quantum circuit of length $n = 4$ implementing the discrete-time reset protocol with reset probability $p$ to the reset state $\ket{\psi_0}$. 
At the end of the circuit the reduced or full density matrix is reconstructed by quantum state tomography (QST). (b) Geometric distribution $p(1-p)^m$ of the age $m$ Eq.~(\ref{eq:age_law})], the number of unitary steps since the most recent reset, for $p = 0.1$; the inset shows the reset-free block of the circuit. (c) Population probabilities of the states $\ket{00}$, $\ket{01}$, $\ket{10}$, and $\ket{11}$ for the product-state reset $\ket{\psi_0} = \ket{00}$, at $\Omega = 1$, $\alpha = 3.2$, and $p = 0.1$, as a function of the total number of steps $n$. The probabilities for $\ket{01}$ and $\ket{10}$ coincide. With a finite reset probability the system reaches a nonequilibrium stationary state (dashed lines); without reset (solid lines) it undergoes coherent oscillations.
}
    \label{fig.1:circuit}
\end{figure}

\section{Model and reset protocol}
\label{sec:model}

{We first study} two interacting qubits described by the transverse-field Ising Hamiltonian,
\begin{equation}
H = -J\,\sigma_1^z\sigma_2^z
    + \frac{\Omega}{2}\left(\sigma_1^x + \sigma_2^x\right),
\label{eq:hamiltonian_2qubit}
\end{equation}
with ferromagnetic coupling $J > 0$ and transverse field $\Omega > 0$. We work in the basis of eigenstates of $\sigma_i^z$, where $i = 1, 2$ labels the two spins; the Hilbert space has dimension 4 and is spanned by $\ket{\uparrow\uparrow}$, $\ket{\uparrow\downarrow}$, $\ket{\downarrow\uparrow}$, and $\ket{\downarrow\downarrow}$, which we also write as $\ket{00}$, $\ket{01}$, $\ket{10}$, and $\ket{11}$ with $\ket{0} \equiv \ket{\uparrow}$ and $\ket{1} \equiv \ket{\downarrow}$, so that $\sigma^z\ket{0} = +\ket{0}$ as in the computational basis of the processor. In this basis the Hamiltonian of Eq.~\eqref{eq:hamiltonian_2qubit} is the $4 \times 4$ matrix
\begin{equation}
H = \begin{pmatrix} -J & \Omega/2 & \Omega/2 & 0 \\
      \Omega/2 & J & 0 & \Omega/2 \\
      \Omega/2 & 0 & J & \Omega/2 \\
      0 & \Omega/2 & \Omega/2 & -J
\end{pmatrix}.
\label{eq:ham_mat}
\end{equation}

This is the minimal interacting setting in which the discrete-time resetting protocol of Wald~\textit{et al.}~\cite{2025_Wald_discrete} produces a stationary state that can be entangled. It is also the minimal such setting in which an analytic solution can be compared directly with data from a superconducting processor and with numerical simulations of the corresponding chain. The extension of Eq.~(\ref{eq:hamiltonian_2qubit}) to a chain of $N$ sites with the same couplings is the standard transverse-field Ising chain, which is studied using exact numerical calculations in Sec.~\ref{sec:exact_chain}.

The two couplings enter the dynamics only through the dimensionless ratio
\begin{equation}
\alpha \equiv \frac{J}{\Omega}\,,
\label{eq:alpha_def}
\end{equation}
and the combination
\begin{equation}
\gamma \equiv \sqrt{\alpha^2 + 1}\,,
\label{eq:gamma_def}
\end{equation}
which arises in diagonalizing (\ref{eq:hamiltonian_2qubit}) and sets the two phase scales of the evolution,
\begin{equation}
\Omega\alpha = J \quad \text{and} \quad \Omega\gamma = \sqrt{J^2 + \Omega^2}.
\label{eq:phase_scales}
\end{equation}

The Hamiltonian of Eq.~(\ref{eq:hamiltonian_2qubit}) has four eigenvalues, $E_{1,2} = \mp\Omega\alpha$ and $E_{3,4} = \mp\Omega\gamma$, derived from its parity block-diagonalization in Sec.~\ref{sec:resonances}. We keep $\Omega$ general in all closed forms: the heatmaps of Sec.~\ref{subsec:heatmap_diagnostics} explore the $(\Omega, J)$ plane, while the hardware experiments of Sec.~\ref{sec:diagnostics}, the chain evaluations of Sec.~\ref{sec:exact_chain}, and the noise analysis of Sec.~\ref{sec:noise} are all performed for $\Omega = 1$, where $\alpha = J$.

At each discrete time step of duration $T$ (Fig.~\ref{fig.1:circuit}), with probability $p \in [0, 1]$ the system is reinitialized to a fixed pure state $\ket{\psi_0}$, and with probability $1 - p$ it evolves under the unitary $U(T) = e^{-iHT}$,
\begin{equation}
\ket{\psi(t+T)}=
\begin{cases}
\ket{\psi_0}, & \text{with prob. } p,\\[4pt]
e^{-iHT}\ket{\psi(t)}, & \text{with prob. } 1-p.
\end{cases}
\label{eq:protocol}
\end{equation}

Throughout this work we set $T = 1$ in units where $J$ and $\Omega$ are measured in inverse time. We consider two reset states in detail:
\begin{enumerate}
 \item the product state $\ket{\psi_0} = \ket{00}$, and
 \item a maximally entangled Bell state $\ket{\psi_0} = (\ket{00} + \ket{11})/\sqrt{2}$, which we refer to as the Bell reset.
\end{enumerate}
The two choices probe complementary aspects of the resetting dynamics. The product-state reset re-initializes the system to a state with no entanglement, and the unitary evolution then generates correlations. The Bell reset re-initializes to a maximally entangled state, on which the unitary evolution acts as a partial disentangler. The qualitative difference between the two reset choices appears in the resonance structure of the stationary state. By a resonance we refer to a curve in the $(\Omega, J)$ plane on which Eq.~(\ref{eq:resonance_intro}) holds for a pair of eigenstates of $H$, so that their relative phase advances by an integer multiple of $2\pi$ at every unitary step. Under this condition the unitarily evolved copies of the reset state that constitute the stationary mixture add with the same phase instead of canceling out. Along such a curve the entanglement diagnostics of the stationary state develop a peak (dip), narrow in the $(\Omega, J)$ plane, visible in Figs.~\ref{fig:heatmap_00} and \ref{fig:heatmap_Bell} as narrow arcs and lines against a smooth background. Which curves show which feature is determined by the reset state, and Sec.~\ref{subsec:selection} derives those rules.

The ensemble-averaged state at time step $n$ satisfies the update rule
\begin{equation}
\bar\rho_{n+1} = (1 - p)\,U \bar\rho_n U^\dagger + p\,\rho_0,
\label{eq:update}
\end{equation}
with $\rho_0 = \ket{\psi_0}\bra{\psi_0}$, where $U(T) = e^{-iHT}$ with $T = 1$ as above. The dynamics has the structure of a renewal process. A reset occurs independently with probability $p$ at each step, so the reset events form a Bernoulli process~\cite{2023_Singh_Bernoulli_reset}, and the state at any step is determined by the number of steps since the most recent reset, geometrically distributed with parameter $p$. Iterating (\ref{eq:update}) from $\bar\rho_0 = \rho_0$ and using the normalization $p\sum_{m=0}^{\infty}(1-p)^m = 1$, the total time $n \to \infty$ limit in Eq.~\eqref{eq:update} yields the stationary state
\begin{equation}
\bar\rho_\infty = p \sum_{m=0}^{\infty}(1-p)^m\,
                  U^m \rho_0 U^{\dagger m},
\label{eq:NESS}
\end{equation} 
which is a convex mixture of unitarily evolved copies of the reset state, weighted by the geometric distribution of reset-free intervals \cite{2021_Perfetto_resetting,2025_Wald_discrete}. This discrete-time renewal sum is the direct analog of the continuous-time density-matrix evolution under Poissonian resetting introduced by Mukherjee~\textit{et al.}~\cite{2018_Mukherjee_reset}, with the geometric age distribution replacing the exponential waiting-time density.

The number of unitary steps $m$ since the most recent reset is the backward recurrence time, or age, of this Bernoulli process~\cite{2001_Godreche_occupation,1971_Feller_probability,2023_Barkai_ergodic}. Its stationary law is the geometric weight
\begin{equation}
\Pr(m) = p\,(1-p)^{m},
\label{eq:age_law}
\end{equation}
with mean $\langle m \rangle = (1-p)/p$. The age includes $m = 0$, the freshly reset state: we read the diagnostics immediately after each step, which fixes the weight of $\rho_0$ at $p$. This is the measurement-time convention set by the protocol of Fig.~\ref{fig.1:circuit}; reading immediately before each step instead would exclude $m = 0$ and reweight the ensemble by $(1-p)^{m-1}$. The stationary state (\ref{eq:NESS}) also satisfies the fixed-point equation
\begin{equation}
\bar\rho_\infty = (1 - p)\,U \bar\rho_\infty U^\dagger + p\,\rho_0.
\label{eq:fixedpoint}
\end{equation}
In addition, the renewal-sum representation makes it explicit that any nonlinear functional of the ensemble (such as a von Neumann entropy or concurrence computed along individual realizations) is in general different from the same functional evaluated on the averaged state $\bar\rho_\infty$, because the convex sum and the nonlinear functional do not commute.

Both entropy diagnostics of this paper start from $\rho_A(m)$, the reduced density matrix of a subsystem $A$ in a realization of age $m$, that is, a realization whose last reset happened $m$ steps earlier. Here $A$ is qubit 1 for two qubits, and half of the chain for the chains of Sec.~\ref{sec:exact_chain}. Section~\ref{subsec:rho_construction} constructs $\rho_A(m)$ for two qubits and the two reset states. The entropy diagnostics apply the same two operations in opposite orders: calculating the von Neumann entropy, and averaging over the age with the weights $p(1-p)^m$ of Eq.~(\ref{eq:age_law}). Averaging first gives the averaged-state entropy, the von Neumann entropy of the averaged reduced state,
\begin{equation}
S(\bar\rho_A^{(\infty)}) = S\!\left[\,
  p \sum_{m=0}^{\infty} (1 - p)^m\,\rho_A(m)\,\right].
\label{eq:S_rhobar}
\end{equation}
Calculating the entropy first gives the trajectory-averaged entropy, the mean von Neumann entropy of the realizations,
\begin{equation}
\bar S_A = p \sum_{m=0}^{\infty} (1 - p)^m\,
           S\!\left[\rho_A(m)\right].
\label{eq:Sbar}
\end{equation}
In addition to these, the third diagnostic we study is the Wootters concurrence $C(\bar\rho_\infty)$ of the averaged two-qubit state~\cite{1998_Wootters_concurrence}, defined in the next section [Eq.~(\ref{eq:C1})]. It measures the entanglement of the stationary mixture itself rather than that of its individual realizations. Section~\ref{sec:diagnostics} calculates all three diagnostics and compares them with measurements on an IBM superconducting processor, where noise acts at every step.

\section{Reduced density matrix and phase matching for the product-state and Bell reset}\label{sec:reduced_density_matrix}
\label{subsec:rho_construction}

{All three diagnostics are nonlinear functions of the density matrix, the two entropies of the reduced state of qubit 1 and the concurrence of the full two-qubit state, so evaluating them requires the explicit construction of $\bar\rho_\infty$ [Eq.~\eqref{eq:NESS}].} 

\subsection{Reduced density matrix}

We perform the construction here for (i) the product-state reset,
\begin{equation}
\ket{\psi_0} = \ket{00},
\label{eq:reset_product}
\end{equation}
and (ii) the Bell reset,
\begin{equation}
\ket{\psi_0} = \ket{\Phi^+} = \frac{\ket{00} + \ket{11}}{\sqrt{2}}.
\label{eq:reset_bell}
\end{equation}
In this two-qubit case, $\rho_A = \mathrm{Tr}_B\,\rho$, and the entries of $\rho_A$ bear a superscript naming the reset state they belong to, since the two resets give different reduced states.

A single realization at age $m$ is the pure state $\rho(m) \equiv U^m \rho_0 U^{\dagger m}$, distinct from the ensemble-averaged $\bar\rho_n$ of Eq.~(\ref{eq:update}), which mixes ages at a fixed time step $n$. The partial trace over qubit $B$ gives the reduced state $\rho_A(m) = \mathrm{Tr}_B\,\rho(m)$, the $2 \times 2$ matrix
\begin{equation}
\rho_A(m) = \begin{pmatrix}
\rho_{\uparrow\uparrow,\uparrow\uparrow}(m) + \rho_{\uparrow\downarrow,\uparrow\downarrow}(m) & \rho_{\uparrow\uparrow,\downarrow\uparrow}(m) + \rho_{\uparrow\downarrow,\downarrow\downarrow}(m) \\
\rho_{\downarrow\uparrow,\uparrow\uparrow}(m) + \rho_{\downarrow\downarrow,\uparrow\downarrow}(m) &
\rho_{\downarrow\uparrow,\downarrow\uparrow}(m) + \rho_{\downarrow\downarrow,\downarrow\downarrow}(m)
\end{pmatrix}\, .
\label{eq:rhoAt}
\end{equation}

For the product-state reset $\ket{00}$, direct calculation of the reduced state of qubit 1 at age $m$, gives
\begin{equation}
\rho_A(m) = \begin{pmatrix}  V(m) & W(m) \\ W^*(m) & 1-V(m) \end{pmatrix},
\label{eq:rho_age}
\end{equation}
with $V(m) = V^{(00)}(m)$ and $W(m) = W^{(00)}(m)$ given by
\begin{align}
V^{(00)}(m) &= \frac{1}{2}\Bigl[1 + \cos(\Omega\alpha m)\cos(\Omega\gamma m) \notag \\
     &\qquad\quad + (\alpha/\gamma)\sin(\Omega\alpha m)\sin(\Omega\gamma m)\Bigr],
        \label{eq:Vm} \\
W^{(00)}(m) &= -\frac{\sin(\Omega\gamma m)\,
        [\alpha\sin(\Omega\gamma m) - i\gamma\cos(\Omega\alpha m)]}{2\gamma^2},
        \label{eq:Wm}
\end{align}
in terms of the ratio $\alpha$ of Eq.~(\ref{eq:alpha_def}) and the combination $\gamma$ of Eq.~(\ref{eq:gamma_def}); the two relevant phases are $\Omega\alpha m = Jm$ and $\Omega\gamma m = \sqrt{J^2+\Omega^2}\,m$, and the superscript $(00)$ marks the product-state reset. The stationary state requires the renewal averages of these entries, $V_\infty \equiv \langle V \rangle = p\sum_{m=0}^{\infty}(1-p)^m V(m)$ and likewise $W_\infty$, obtained by substituting (\ref{eq:Vm}) and (\ref{eq:Wm}) into the geometric sums prescribed by (\ref{eq:NESS}). Since $V(m)$ and $W(m)$ are sums of sinusoids in $m$, these averages can be written compactly with one function, the complex geometric kernel
\begin{equation}
\mathcal{G}(w) \equiv p\sum_{m=0}^{\infty}(1-p)^m e^{imw} = \frac{p}{1-(1-p)e^{iw}}\,,
\label{eq:kernel_complex}
\end{equation}
the resummed geometric series of the reset-free phase $e^{iw}$ weighted by the geometric age distribution $p(1-p)^m$ of Eq.~(\ref{eq:age_law}). Its real and imaginary parts are the two geometric kernels
\begin{eqnarray}
G_c(w) &\equiv& \mathrm{Re}\,\mathcal{G}(w) = \frac{p\,[1-(1-p)\cos w]}{D(w)}\,,
\nonumber\\[1mm]
G_s(w) &\equiv& \mathrm{Im}\,\mathcal{G}(w) = \frac{p(1-p)\sin w}{D(w)}\,,
\label{eq:kernels}
\end{eqnarray}
whose common denominator
\begin{equation}
D(w) = \bigl|\,1 - (1-p)e^{iw}\,\bigr|^2 = 2(p-1)\cos w + (p-2)p + 2
\label{eq:Dw}
\end{equation}
takes its minimum value $p^2$ when $w \in 2\pi\mathbb{Z}$, where $G_c = 1$ and $G_s = 0$. Evaluating the geometric sums term by term with these kernels~\cite{2024_Wolfram_mathematica} gives
\begin{widetext}
\begin{align}
V^{(00)}_\infty &= \frac{1}{2}
  + \frac{1}{4}\Bigl(1+\frac{\alpha}{\gamma}\Bigr)\,G_c\bigl(\Omega(\alpha-\gamma)\bigr)
  + \frac{1}{4}\Bigl(1-\frac{\alpha}{\gamma}\Bigr)\,G_c\bigl(\Omega(\alpha+\gamma)\bigr),
\label{eq:Vinf} \\[4pt]
W^{(00)}_\infty &= -\frac{\alpha}{4\gamma^2}\bigl[\,1 - G_c(2\Omega\gamma)\,\bigr]
  + \frac{i}{4\gamma}\bigl[\,G_s\bigl(\Omega(\gamma+\alpha)\bigr) + G_s\bigl(\Omega(\gamma-\alpha)\bigr)\,\bigr].
\label{eq:Winf}
\end{align}
\end{widetext}

For the Bell reset, Eq.~\eqref{eq:rho_age} holds with $V(m) = V^{(\Phi)}(m)$ and $W(m) = W^{(\Phi)}(m)$,
\begin{align}
V^{(\Phi)}(m) = \frac{1}{2},\,\,\,
W^{(\Phi)}(m) = -\frac{\alpha  \sin ^2\left(\Omega\gamma  \,m\right)}{\gamma^2}\,,
        \label{eq:Wmbell}
\end{align}
where the superscript $(\Phi)$ marks the Bell reset. Substituting (\ref{eq:Wmbell}) into the geometric sums prescribed by (\ref{eq:NESS}) yields
\begin{eqnarray}
V^{(\Phi)}_{\infty} =  \frac{1}{2}\,,\,
W^{(\Phi)}_{\infty} = -\frac{\alpha}{2\gamma^2}\bigl[\,1 - G_c(2\Omega\gamma)\,\bigr]\,,
\label{eq:Winfbell}
\end{eqnarray}
which is the same kernel that produces the resonance in (\ref{eq:Winf}), with twice its weight because the Bell reset places all of its coherence in the parity-even sector.

For either reset, the reduced single-qubit stationary state has the same form,
\begin{equation}
\bar\rho_A^{(\infty)} = \begin{pmatrix}  V_\infty & W_\infty \\ W_\infty^* & 1-V_\infty \end{pmatrix},
\label{eq:rhoA_inf}
\end{equation}
with $V_\infty$ and $W_\infty$ the pair belonging to the reset state under consideration, Eqs.~(\ref{eq:Vinf}) and (\ref{eq:Winf}) for the product-state reset and Eq.~(\ref{eq:Winfbell}) for the Bell reset, with eigenvalues
\begin{equation}
\lambda_\pm^{(\infty)}
  = \frac{1}{2}\!\left[1 \pm
    \sqrt{(2V_\infty - 1)^2 + 4|W_\infty|^2}\right].
\label{eq:eigenvalues}
\end{equation}
The von Neumann entropy of the reduced state follows from the binary entropy function $h$:
\begin{eqnarray}
\label{eq:S_avgrho}
S(\bar\rho_A^{(\infty)})
  &=& h\!\left(\lambda_+^{(\infty)}\right)\nonumber \quad \text{with}\\
h(x) &\equiv& -x\log_2 x - (1 - x)\log_2(1 - x).
\end{eqnarray}
The two eigenvalues sum to unity, $\lambda_+^{(\infty)} + \lambda_-^{(\infty)} = 1$, since $\bar\rho_A^{(\infty)}$ has unit trace, and $h(x) = h(1-x)$, so either eigenvalue may be inserted and $\lambda_+^{(\infty)}$ is chosen only for definiteness. Entropies are in bits throughout.

The Wootters concurrence of a two-qubit state $\rho$ is
\begin{equation}
C(\rho) = \max (0, \mu_1 - \mu_2 - \mu_3 - \mu_4)\, ,
\label{eq:C1}
\end{equation}
where $\mu_1 \geq \mu_2 \geq \mu_3 \geq \mu_4$ are the eigenvalues of
\begin{equation}
\mathcal{R} =  \sqrt{\sqrt{\rho}\, \tilde{\rho} \, \sqrt{\rho}} \, ,
\label{eq:C2}
\end{equation}
with the spin-flipped state
\begin{equation}
\tilde{\rho} = (\sigma^y \otimes \sigma^y) \, \rho^* \,  (\sigma^y \otimes \sigma^y) \, ,
\label{eq:C3}
\end{equation}
where $\sigma^{x}$, $\sigma^{y}$, and $\sigma^{z}$ are the Pauli matrices and $\rho^*$ is the complex conjugate of $\rho$ in the $\sigma^{z}$ computational basis. The concurrence takes values in $[0, 1]$ and equals $1$ for a maximally entangled state. The diagnostic used below is $C(\bar\rho_\infty)$, the concurrence of the averaged two-qubit stationary state.

\subsection{Spectral resonances and phase matching}
\label{sec:resonances}

In this subsection we derive the phase-matching condition that governs the two-qubit stationary state and show, from the Hamiltonian's parity symmetry, that it collapses to four distinct loci in the $(\Omega, J)$ plane.

The dependence on $\Omega$ and $J$ of the closed forms (\ref{eq:Vinf}) and (\ref{eq:Winf}) enters only through the geometric kernels (\ref{eq:kernels}), and those in turn only through the denominator $D$ of Eq.~(\ref{eq:Dw}), which is smallest exactly where a phase is an integer multiple of $2\pi$. In terms of the Hamiltonian, the condition follows from evaluating (\ref{eq:NESS}) in the eigenbasis of $H$. Inserting resolutions of the identity into the renewal sum and performing the geometric series term by term gives
\begin{equation}
\bra{E_i}\bar\rho_\infty\ket{E_j}
  = \frac{p\,\bra{E_i}\rho_0\ket{E_j}}
         {1 - (1 - p)\,e^{-i(E_i - E_j)T}},
\label{eq:eigenbasis}
\end{equation}
where $\{\ket{E_k}\}$ are the eigenstates of $H$ with eigenvalues $E_k$ [Eq.~\eqref{eq:hamiltonian_2qubit}]. The factor multiplying $\bra{E_i}\rho_0\ket{E_j}$ in Eq.~(\ref{eq:eigenbasis}) is the geometric kernel of Eq.~(\ref{eq:kernel_complex}), evaluated at the phase $(E_j - E_i)T$ that the two levels acquire relative to each other in one step. Each matrix element of the stationary state in the energy eigenbasis is therefore the corresponding element of the reset state multiplied by this kernel, $\bra{E_i}\bar\rho_\infty\ket{E_j} = \mathcal{G}\bigl((E_j - E_i)T\bigr)\bra{E_i}\rho_0\ket{E_j}$.

When the phase difference satisfies $(E_i - E_j)T \in 2\pi\mathbb{Z}$, the condition stated in Eq.~(\ref{eq:resonance_intro}), the denominator equals $p$, the factor $p$ in the numerator cancels, and $\bra{E_i}\bar\rho_\infty\ket{E_j} = \bra{E_i}\rho_0\ket{E_j}$: the coherence of the reset state survives the averaging in full, at every reset probability. Away from that condition the denominator is of order unity and the same element is suppressed by a factor of order $p$, so the reset probability sets how far the resonance stands above the background but not where it lies. This is the spectral resonance mechanism identified in a general setting by Wald \textit{et al.} \cite{2025_Wald_discrete} for arbitrary discrete-time reset protocols. The element also vanishes for every $p$ when $\bra{E_i}\rho_0\ket{E_j} = 0$, whether or not the phase condition is met; the phase-matching condition alone does not produce a resonance, since the reset state must also overlap with both eigenstates for their coherence to appear in the stationary state. The Hamiltonian therefore fixes where the resonances lie and the reset state fixes which of them survive, and neither is a function of $p$.

The four eigenenergies of the Hamiltonian in Eq.~(\ref{eq:hamiltonian_2qubit}) generate six pairwise phase-matching conditions, worked out in the remainder of this section; the selection rule that follows from the overlap prefactor is described in Sec.~\ref{subsec:selection}. The Hamiltonian of Eq.~(\ref{eq:hamiltonian_2qubit}) commutes with the global spin-flip operator $P = \sigma_1^x \sigma_2^x$, and the Hilbert space decomposes accordingly into two parity sectors of dimension two. The transverse-field term commutes with $P$ trivially, and the Ising term $-J\sigma_1^z \sigma_2^z$ commutes with $P$ because each on-site anticommutation $\{\sigma_i^z, \sigma_i^x\} = 0$ contributes a sign that is canceled by the corresponding sign on the other qubit.

Within each sector the Hamiltonian reduces to that of a single pseudospin. Writing $\sigma^x$ and $\sigma^z$ without a site index for the Pauli matrices acting on a two-dimensional sector in the ordered basis given for it, the parity-even sector is spanned by $\ket{\Phi^+} = (\ket{00} + \ket{11})/\sqrt 2$ and $\ket{\Psi^+} = (\ket{01} + \ket{10})/\sqrt 2$, in which
\begin{equation}
H^{(+)} = -J\,\sigma^z + \Omega\,\sigma^x ,
\label{eq:H_even}
\end{equation}
with eigenvalues $E_{3,4} = \mp\sqrt{J^2 + \Omega^2} = \mp\,\Omega\gamma$ and eigenstates
\begin{align}
\ket{E_3} &= \cos(\theta/2)\ket{\Phi^+} - \sin(\theta/2)\ket{\Psi^+},
\label{eq:eigenstates_plus_E3}
\\
\ket{E_4} &= \sin(\theta/2)\ket{\Phi^+} + \cos(\theta/2)\ket{\Psi^+},
\label{eq:eigenstates_plus}
\end{align}
where the mixing angle is fixed by $\tan\theta = \Omega/J$.

The parity-odd sector is spanned by $\ket{\Phi^-} = (\ket{00} - \ket{11})/\sqrt 2$ and $\ket{\Psi^-} = (\ket{01} - \ket{10})/\sqrt 2$, on both of which the transverse field vanishes, as direct action of $\sigma_1^x + \sigma_2^x$ shows. In the same notation
\begin{equation}
H^{(-)} = -J\,\sigma^z ,
\label{eq:H_odd}
\end{equation}
with eigenvalues $E_{1,2} = \mp J = \mp\,\Omega\alpha$ and eigenstates $\ket{E_1} = \ket{\Phi^-}$ and $\ket{E_2} = \ket{\Psi^-}$. The two sectors differ by the transverse-field term and by nothing else: $\Omega$ splits the even doublet to $\pm\sqrt{J^2 + \Omega^2}$ and leaves the odd doublet at $\pm J$. The four eigenvalues therefore involve the two energy scales $J$ and $\sqrt{J^2 + \Omega^2}$, which are $\Omega\alpha$ and $\Omega\gamma$ in the notation of Sec.~\ref{sec:model}.

Parity constrains the stationary state through the numerator of Eq.~(\ref{eq:eigenbasis}). The eigenstates bear the parity eigenvalues $P\ket{E_{1,2}} = -\ket{E_{1,2}}$ and $P\ket{E_{3,4}} = +\ket{E_{3,4}}$, while the prefactor $\bra{E_i}\rho_0\ket{E_j} = \braket{E_i|\psi_0}\braket{\psi_0|E_j}$ is fixed by the reset state alone. A reset state of definite parity gives a $\rho_0$ that commutes with $P$, all opposite-parity prefactors vanish, and only same-parity pairs can resonate; this is the condition under which resonances are restricted to eigenstates of equal parity, as in the monitored quantum walk of Singh~\textit{et al.}~\cite{2026_Singh_splitting}.

The reset protocol admits reset states of mixed parity, for which the restriction does not hold: $\ket{\Phi^+}$ has $P = +1$ and commutes with $P$, whereas $\ket{00} = (\ket{\Phi^+} + \ket{\Phi^-})/\sqrt 2$ is an equal superposition of the two sectors and contains nonzero cross-parity coherences. Which pairs resonate is therefore a property of the reset state and not of the Hamiltonian, and Sec.~\ref{subsec:selection} works out the two cases.

Applied to the six pairs of eigenenergies, Eq.~(\ref{eq:resonance_intro}) leaves four distinct loci. A pair and its reverse give the same condition, and the cross-parity pairs coincide two by two, since $E_1 - E_3 = -(E_2 - E_4)$ and $E_2 - E_3 = -(E_1 - E_4)$. For a positive integer $k$, the two same-parity pairs, $(E_1, E_2)$ in the odd sector and $(E_3, E_4)$ in the even sector, give
\begin{align}
2J\,T &= 2\pi k, \label{eq:cond_1} \\
2\sqrt{J^2 + \Omega^2}\,T &= 2\pi k, \label{eq:cond_2}
\end{align}
and the four cross-parity pairs, $(E_1, E_3)$ and $(E_2, E_4)$ for the first condition and $(E_2, E_3)$ and $(E_1, E_4)$ for the second, give
\begin{align}
\bigl(\sqrt{J^2 + \Omega^2} - J\bigr)\,T &= 2\pi k, \label{eq:cond_3} \\
\bigl(\sqrt{J^2 + \Omega^2} + J\bigr)\,T &= 2\pi k. \label{eq:cond_4}
\end{align}
Each left-hand side is an energy difference $|E_i - E_j|$ times $T$, and each is positive, so $k = 0$ does not occur. The loci are written in $J$ and $\Omega$ rather than in $\alpha$ and $\gamma$ because they are curves in the plane of Figs.~\ref{fig:heatmap_00} and \ref{fig:heatmap_Bell}.

With $T = 1$, as throughout, condition (\ref{eq:cond_1}) describes the lines $J = \pi k$, independent of $\Omega$, which run horizontally in the $(\Omega, J)$ plane of Figs.~\ref{fig:heatmap_00} and \ref{fig:heatmap_Bell}. Condition (\ref{eq:cond_2}) describes quarter-circles centered at the origin with radii equal to integer multiples of $\pi$. Condition (\ref{eq:cond_3}) rearranges to $\Omega^2 = 4\pi k J + 4\pi^2 k^2$, describing a parabola opening towards positive $J$ that crosses $\Omega = 0$ at $J = -\pi k$, outside the physical region. It enters the first quadrant only for $\Omega > 2\pi k$. Condition (\ref{eq:cond_4}) rearranges to $J = \pi k - \Omega^2/(4\pi k)$, a parabola opening towards negative $J$ with vertex at $(J, \Omega) = (\pi k, 0)$.

At the typical operating point $\Omega = 1$ used in the hardware measurements, the smallest values of $J$ at which each condition is satisfied are $J = \pi \approx 3.142$ from (\ref{eq:cond_1}), $J = \sqrt{\pi^2 - 1} \approx 2.978$ from (\ref{eq:cond_2}), no real solution from (\ref{eq:cond_3}) for $k = 1$, and $J = \pi - 1/(4\pi) \approx 3.062$ from (\ref{eq:cond_4}). These values cluster within a narrow window $|\Delta J| \approx 0.16$ at $\Omega = 1$, and depending on the heatmap resolution they may be resolved as distinct features or appear as a single broad feature.

The overlap of the reset state with the eigenstates selects which of these curves appear in the averaged state $\bar\rho_\infty$ (Sec.~\ref{subsec:selection}). The trajectory-averaged entropy is evaluated before the reset average is taken, and it can show further curves that $\bar\rho_\infty$ does not contain (Sec.~\ref{sec:diagnostics}). The heatmaps of Figs.~\ref{fig:heatmap_00} and \ref{fig:heatmap_Bell} display both effects (Sec.~\ref{subsec:heatmap_diagnostics}).

\subsubsection*{Remark on the continuous time limit.}\par

Restoring $T$ and taking $T \to 0$ at a fixed reset rate $r \equiv p/T$,
the age becomes a continuous time $t = mT$ and Eq.~\eqref{eq:NESS} reduces to
$\bar\rho_\infty = r\int_0^\infty dt\, e^{-rt} e^{-iHt}\rho_0\, e^{iHt}$
\cite{2018_Mukherjee_reset}. Equation~\eqref{eq:eigenbasis} then becomes
\begin{equation}
\bra{E_i}\bar\rho_\infty\ket{E_j}
  = \frac{r\,\bra{E_i}\rho_0\ket{E_j}}{r + i(E_i - E_j)} .
\label{eq:eigenbasis_continuum}
\end{equation}
The denominator of Eq.~\eqref{eq:eigenbasis} is $2\pi$-periodic in
$(E_i - E_j)T$ because the age takes only integer multiples of $T$, so two
levels whose relative phase advances by $2\pi k$ per step act as if degenerate.
Equation~\eqref{eq:eigenbasis_continuum} has no such periodicity: its magnitude is
largest at $E_i = E_j$, and a coherence survives only between levels with
$|E_i - E_j| \lesssim r$. The resonances at $E_i - E_j = 2\pi k/T$, $k \neq 0$,
move to infinite energy as $T \to 0$, so the loci
(\ref{eq:cond_1}) to (\ref{eq:cond_4}) have no continuous-time counterpart, and the
resonance arcs of Figs.~\ref{fig:heatmap_00} and \ref{fig:heatmap_Bell} disappear.
This is consistent with Ref.~\cite{2025_Wald_discrete}, which identifies the
resonances as a feature of discrete time.

\subsection{Selection by the reset state}
\label{subsec:selection}

Equation (\ref{eq:eigenbasis}) shows that the contribution of any phase-matching pair to the stationary state is multiplied by the prefactor $\bra{E_i}\rho_0\ket{E_j}$. This prefactor vanishes whenever the reset state $\ket{\psi_0}$ has zero overlap with either $\ket{E_i}$ or $\ket{E_j}$, and the corresponding resonance is then kinematically suppressed regardless of whether the phase-matching condition is satisfied.

Since $\rho_0 = \ket{\psi_0}\bra{\psi_0}$ is a pure state, the prefactor factorizes as $\bra{E_i}\rho_0\ket{E_j} = c_i c_j^{*}$, where $c_k \equiv \braket{E_k|\psi_0}$ is the overlap of the reset state with the $k$-th eigenstate, and Eq.~(\ref{eq:eigenbasis}) becomes
\begin{equation}
\bra{E_i}\bar\rho_\infty\ket{E_j}
  = \frac{p\, c_i\, c_j^{*}}{1 - (1 - p)\,e^{-i(E_i - E_j)T}} .
\label{eq:eigenbasis_overlap}
\end{equation}
The pair $(i, j)$ contributes to the stationary state if and only if $c_i c_j^{*} \neq 0$, so the entire selection rule is fixed by the overlap vector. Writing it in the order $(E_1, E_2, E_3, E_4)$ and reading the eigenstates from (\ref{eq:eigenstates_plus_E3}) and (\ref{eq:eigenstates_plus}),
\begin{align}
c^{(00)} &= \tfrac{1}{\sqrt 2}\bigl(1,\ 0,\ \cos(\theta/2),\ \sin(\theta/2)\bigr),
\label{eq:overlap_00} \\
c^{(\Phi^+)} &= \bigl(0,\ 0,\ \cos(\theta/2),\ \sin(\theta/2)\bigr).
\label{eq:overlap_Bell}
\end{align}
Both reset states have $c_2 = 0$, so the pair $(1, 2)$ never contributes and condition (\ref{eq:cond_1}), the lines $J = \pi k$, is absent from $\bar\rho_\infty$ for either reset. The product-state reset has $c_1 \neq 0$, and its supported pairs $(3, 4)$, $(1, 3)$, and $(1, 4)$ realize conditions (\ref{eq:cond_2}), (\ref{eq:cond_3}), and (\ref{eq:cond_4}) respectively: the odd-sector component $\ket{E_1}$ that $\ket{00}$ contains is what activates the two cross-parity loci. The Bell reset has $c_1 = c_2 = 0$, is supported on the parity-even sector alone, and contributes through the single pair $(3, 4)$, so condition (\ref{eq:cond_2}) is the only locus it realizes.

\section{Entropy and concurrence: from theory to quantum hardware}
\label{sec:diagnostics}

The selection rule of Sec.~\ref{subsec:selection} constrains the matrix elements of $\bar\rho_\infty$, and with them every quantity computed from the averaged state, including the averaged-state entropy $S(\bar\rho_A^{(\infty)})$ of Eq.~(\ref{eq:S_avgrho}). The trajectory-averaged entropy $\bar S_A$ is not such a quantity: the von Neumann entropy of each reset realization is computed first, and the average over realizations is taken afterwards. The two diagnostics therefore differ only in the order of two operations, calculating the von Neumann entropy and averaging over reset histories, and because the von Neumann entropy is concave this order is important. For a concave function, the average of the function never exceeds the function of the average.

In this section we study both entropy diagnostics, show that their difference is an observable that cannot be computed from $\bar\rho_\infty$ alone, and compare both, together with the concurrence, with the measurement on \mbox{ibm\_pittsburgh}~\cite{2026_IBM_platform,2026_Murauer_NESS}.

\subsection{The two averaging protocols for entropy}
\label{subsec:two_protocols}

The von Neumann entropy of $\rho_A(m)$, the reduced state of the age-$m$ realization defined in Sec.~\ref{subsec:rho_construction}, measures the entanglement of that realization. In the stationary ensemble, the age is drawn from the geometric law of Eq.~(\ref{eq:age_law}), and the total step count $n$ does not enter. Since the reset breaks detailed balance, this stationary ensemble is a nonequilibrium steady state.

In the averaged-state entropy of Eq.~(\ref{eq:S_rhobar}) the average acts linearly on the reduced states, and the entropy of the averaged state is calculated once. Experimentally, $S(\bar\rho_A^{(\infty)})$ follows from the ensemble-averaged tomographic reconstruction, and its closed form is Eq.~(\ref{eq:S_avgrho}). In the trajectory-averaged entropy the entropy of every realization is calculated directly and separately. Experimentally, $\bar S_A$ requires the tomographic reconstruction to be performed realization by realization, as described in Sec.~\ref{subsec:experimental_significance}.

The two expressions differ because the entropy functional does not commute with the convex average. Concavity of the von Neumann entropy gives the Jensen inequality~\cite{2010_Nielsen_qcqi, 1978_Wehrl_entropy}
\begin{equation}
S(\bar\rho_A^{(\infty)}) \geq \bar S_A,
\label{eq:jensen}
\end{equation}
with equality if and only if every reduced state $\rho_A(m)$ with nonzero weight is the same state, whatever entropy that state has. At $p = 1$ only the $m = 0$ term contributes, and both diagnostics equal the entropy of the reset state: 1 bit for the Bell reset, and 0 for the product-state reset, which is unentangled. Equality also holds on parameter loci where $\rho_A(m)$ is independent of $m$, since the ensemble is then a single state at every $p$; the Bell reset at $\Omega\gamma = \pi$ is such a locus, because $U\ket{\Phi^+} = -\ket{\Phi^+}$ there and every realization has the same reduced state, which is maximally mixed and gives 1 bit. Equality therefore reports that the ensemble has collapsed to one reduced state, whatever its entropy. Away from these cases the inequality is strict, and the difference
\begin{equation}
\Delta S(\alpha, p) \equiv S(\bar\rho_A^{(\infty)}) - \bar S_A \geq 0,
\label{eq:DeltaS_def}
\end{equation}
is the Jensen gap of the von Neumann entropy, the amount by which the concavity inequality is strict. Since it is the only Jensen gap in this paper, we call it the Jensen gap for short. It is an observable of the reset ensemble that cannot be obtained from $\bar\rho_\infty$, because different ensembles of reduced states can average to the same $\bar\rho_A$.

With the age weights $w_m = p(1-p)^m$ of Eq.~(\ref{eq:age_law}), Eq.~(\ref{eq:DeltaS_def}) reads
\begin{align}
\Delta S &= S\Bigl(\sum_{m} w_m\,\rho_A(m)\Bigr) - \sum_{m} w_m\, S[\rho_A(m)] \nonumber \\
&= \chi\bigl(\{w_m, \rho_A(m)\}\bigr),
\label{eq:holevo}
\end{align}
which is by definition the Holevo information $\chi$ of the set of reduced states $\rho_A(m)$, each occurring with the probability $w_m$ of its age~\cite{1973_Holevo_bound}. Holevo introduced this quantity to bound how much classical information a measurement can extract from such a set of states. In quantum information it is also called the quantum Jensen-Shannon divergence of the set~\cite{2009_Briet_jensen}. A Jensen gap can be defined for any concave function. For the von Neumann entropy over a set of states, each with its probability, it is exactly the Holevo information, so here the Jensen gap and the Holevo information are two names for one quantity. We use the name Jensen gap throughout, because Eq.~(\ref{eq:DeltaS_def}) defines it by the order of averaging, and the name Holevo information only where its bound on measurements is used. Suppose one run of the circuit has lasted many time steps $n$, long enough to reach the stationary state, and only subsystem $A$ is measured at its end, with the aim of inferring the age $m$, which is the number of steps since the last reset. If the age $m$ were known, $A$ would be in the state $\rho_A(m)$, and its entropy averaged over the ages is $\bar S_A$. If the age is not known, $A$ is in the averaged state $\bar\rho_A$, whose entropy $S(\bar\rho_A)$ is at least as large, Eq.~(\ref{eq:jensen}). Their difference $\Delta S$ is therefore the uncertainty about $A$ that comes only from not knowing the age; it is the quantum mutual information between $A$ and a classical record of the age, in the joint state $\sum_m w_m \ket{m}\bra{m} \otimes \rho_A(m)$. By Holevo's theorem, no measurement on $A$ alone yields more than $\Delta S$ bits of information about the age~\cite{1973_Holevo_bound}. A gap $\Delta S = 0$ means that every age leaves $A$ in the same state, so $A$ holds no information about the particular age $m$; a gap of $2$ bits allows at most as much information as identifying which of four equally likely groups the age $m$ falls in. Both terms of Eq.~(\ref{eq:holevo}) are obtained by tomography, so $\Delta S$ is measurable; for the Bell reset it is the difference between panels (b) and (a) of Fig.~\ref{fig:S_Sbar_DeltaS}\monalisa, and panel (c) of Figs.~\ref{fig:heatmap_00} and \ref{fig:heatmap_Bell} shows it across the whole $(\Omega, J)$ plane for both reset states.

\subsection{Closed form for the trajectory-averaged entropy}
\label{subsec:Sbar_closedform}

The reduced state $\rho_A(m)$ is the reduction of a two-qubit pure state, so its spectrum is determined by the concurrence of that pure state: the more entangled the two-qubit state, the more mixed the reduced state of either qubit. Writing $\ket{\psi(m)} = U^m \ket{00} = \sum_{ab} c_{ab}(m) \ket{ab}$ for the product-state reset, with coefficients $c_{ab}(m)$ given in Appendix~\ref{app:trajectory_avg_entropy}, the pure-state concurrence at age $m$ is
\begin{equation}
C_m = 2\,|c_{00}(m)\,c_{11}(m) - c_{01}(m)\,c_{10}(m)|,
\label{eq:Cm}
\end{equation}
and the eigenvalues of $\rho_A(m)$ are $\lambda_\pm^{(m)} = (1 \pm \sqrt{1 - C_m^2})/2$. The von Neumann entropy of the single realization at age $m$ is therefore
\begin{equation}
S^{(m)} = h\!\left(\frac{1 + \sqrt{1 - C_m^2}}{2}\right),
\label{eq:Sm}
\end{equation}
with $h(x)$ the binary entropy function of Eq.~(\ref{eq:S_avgrho}), and substitution into (\ref{eq:Sbar}) gives
\begin{equation}
\bar S_A = p \sum_{m=0}^{\infty} (1 - p)^m\,
          h\!\left(\frac{1 + \sqrt{1 - C_m^2}}{2}\right),
\label{eq:Sbar_closedform}
\end{equation}
the counterpart of Eq.~(\ref{eq:S_avgrho}) for the trajectory-averaged diagnostic. The derivation of $C_m$ and the analogous construction for the Bell reset are collected in Appendix~\ref{app:trajectory_avg_entropy}.

Equation (\ref{eq:Sbar_closedform}) does not simplify further because the binary entropy is a nonlinear function of its argument. The sum converges geometrically and is evaluated by truncation at $m_{\max} \sim 10/p$, at which the tail weight $(1-p)^{m_{\max}}$ is of order $e^{-10}$ and is negligible at any experimental or numerical precision of interest; at $p = 0.1$ this requires approximately one hundred terms, and at $p = 0.9$ five terms suffice.

\subsection{The Jensen gap as an independent observable}
\label{subsec:DeltaS}

At fixed $\alpha$ the gap vanishes at $p = 1$ and approaches a finite value as $p \to 0$. At $p = 1$ every realization is the reset state and the ensemble contains one state. As $p$ decreases, longer reset-free evolutions contribute more probability weight, so reduced states of larger age $m$ keep larger weights $p(1-p)^m$, and the gap grows, passing through a maximum at an intermediate reset probability $p^*(\alpha)$. In the limit $p \to 0$ the geometric weights flatten, and both diagnostics become long-time quantities of the reset-free evolution, the sequence of states $U^m\ket{\psi_0}$. The averaged-state entropy approaches the entropy of the time-averaged reduced state, and the trajectory-averaged entropy approaches the time average of the single-realization entropies. The gap approaches their difference, which is positive unless $\rho_A(m)$ is independent of $m$. At $\alpha = 1$ and $\Omega = 1$ the product-state reset gap rises from zero at $p = 1$ to a maximum of $0.50$ at $p^* = 0.13$ and approaches $0.47$ as $p \to 0$; the limit is derived, and the finite-$p$ values are evaluated from Eqs.~(\ref{eq:S_avgrho}) and (\ref{eq:Sbar_closedform}), in Appendix~\ref{app:jensen_limit}.

\begin{figure*}
\centering
\includegraphics[width=0.8\linewidth]{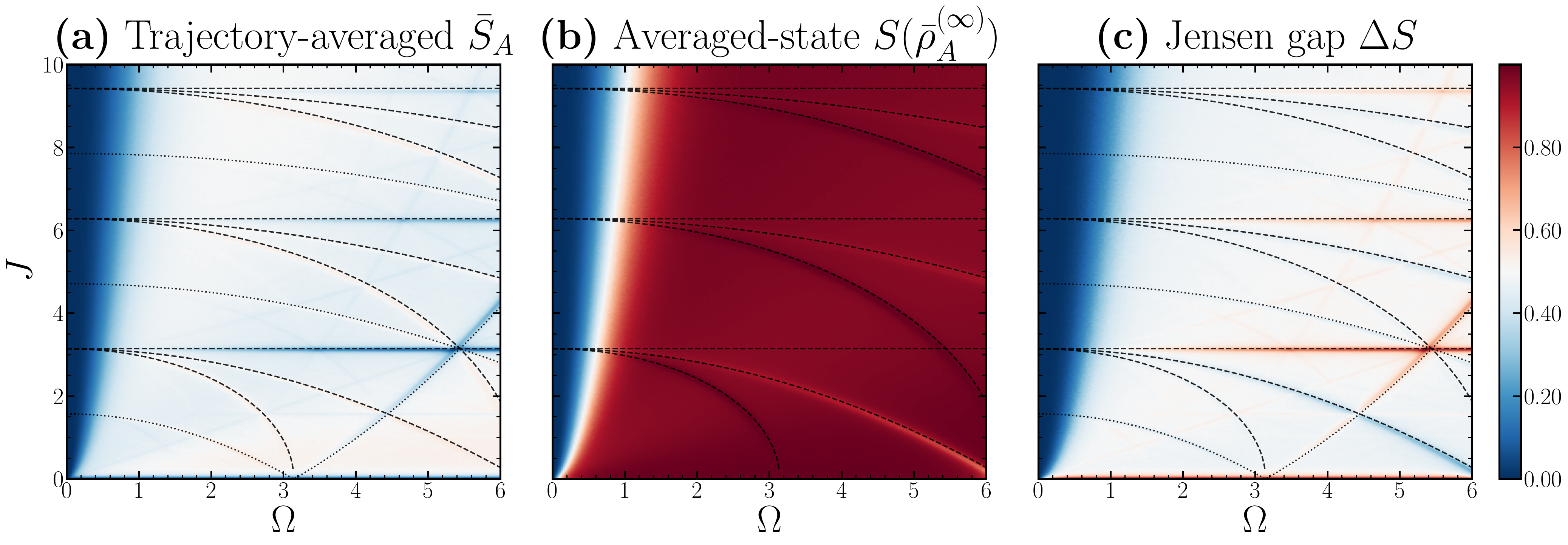}
\caption{Entropy diagnostics of the product-state reset stationary state in the $(\Omega, J)$ plane at $p = 0.1$. (a) Trajectory-averaged entropy $\bar S_A$, Eq.~(\ref{eq:Sbar}), evaluated by trajectory sampling. (b) Averaged-state entropy $S(\bar\rho_A^{(\infty)})$, Eq.~(\ref{eq:S_rhobar}), from the closed form (\ref{eq:S_avgrho}). (c) Jensen gap $\Delta S$, Eq.~(\ref{eq:DeltaS_def}), the difference between the two entropy evaluations. All three panels, and those of Fig.~\ref{fig:heatmap_Bell}, share one colour scale, from 0 (blue) through 0.5 (white) to 1 bit (red). Both entropies are below 0.17 bits at small $\Omega$, where the state is near a Hamiltonian eigenstate, and rise across a crossover whose midpoint lies at $(\sqrt{J^2 + \Omega^2} - J)\,T = 0.75\,p$. Their medians over the plotted plane are 0.46 bits in (a) and 0.96 bits in (b), and the median gap in (c) is 0.49 bits. Dashed black lines mark the phase-matching loci of Eqs.~(\ref{eq:cond_1}) to (\ref{eq:cond_4}), on which a phase difference $(E_i - E_j)T$ is an even multiple of $\pi$. Dotted black lines in (a) and (c) mark the antiperiodic cross-parity loci $\sqrt{J^2 + \Omega^2} \pm J = (2k+1)\pi$, on which it is an odd multiple. In (b) the entropy is raised on the circles of condition (\ref{eq:cond_2}) and lowered along the parabolas of condition (\ref{eq:cond_4}), and the lines $J = \pi k$ leave no feature, as the selection rule of Sec.~\ref{subsec:selection} requires. In (a), $\bar S_A$ is lowered along $J = \pi k$ and along $\sqrt{J^2 + \Omega^2} - J = \pi$, both through the phase collapse of Sec.~\ref{subsec:DeltaS}, and slightly raised along $\sqrt{J^2 + \Omega^2} + J = (2k+1)\pi$. The gap in (c) is accordingly raised (red) along the first two and lowered (blue) along the parabolas. Entropies are in bits.
}
\label{fig:heatmap_00}
\end{figure*}

\begin{figure*}
\centering
\includegraphics[width=0.8\linewidth]{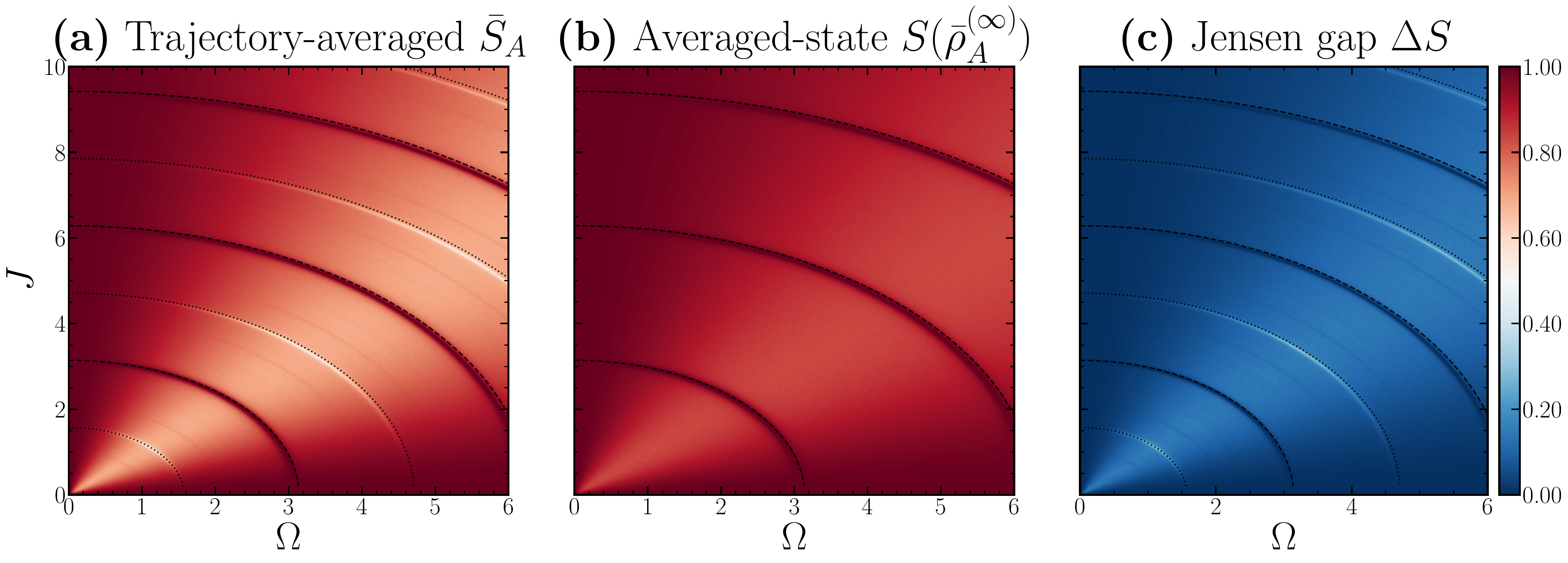}
\caption{Entropy diagnostics of the Bell-reset stationary state in the $(\Omega, J)$ plane at $p = 0.1$. Away from the arcs both entropies depend on $\alpha = J/\Omega$ alone, up to corrections of order $p$. They equal 1 bit along both axes, are lowest near $J = \Omega$, and differ by less than 0.14 bits over 90 percent of the plotted plane. (a) Trajectory-averaged entropy $\bar S_A$, Eq.~(\ref{eq:Sbar}), evaluated by trajectory sampling. The dark arcs of panel (b) recur, and $\bar S_A$ is in addition lowered (light to blue) on the half-integer circles $\sqrt{J^2 + \Omega^2} = (k + \tfrac{1}{2})\pi$. On these circles the realizations alternate between two pure states, and the stationary state has rank 2. No feature occurs on the lines $J = \pi k$. (b) Averaged-state entropy $S(\bar\rho_A^{(\infty)})$, Eq.~(\ref{eq:S_rhobar}). Its only features are the dark arcs on the circles $\sqrt{J^2 + \Omega^2} = \pi k$ of condition (\ref{eq:cond_2}), as the selection rule of Sec.~\ref{subsec:selection} describes. On these circles the stationary state is the pure Bell state, and the entropy equals 1 bit. Dashed black lines trace the circles of condition (\ref{eq:cond_2}), and dotted black lines the half-integer circles in (a). Entropies are in bits.
}
\label{fig:heatmap_Bell}
\end{figure*}

The two entropy diagnostics also resolve resonances through different mechanisms. A feature of $S(\bar\rho_A^{(\infty)})$ requires a resonant matrix element of $\bar\rho_\infty$: a pair of eigenstates, both supported by the reset state, with $(E_i - E_j)T \in 2\pi\mathbb{Z}$, as in Sec.~\ref{subsec:selection}. The trajectory-averaged entropy is a nonlinear function of the trajectory, and it develops features wherever the phases accumulated by a realization repeat periodically with its age $m$, whether or not any supported pair of eigenstates is phase matched. For the Bell reset the trajectory accumulates the single phase $2\Omega\gamma m$: at $2\Omega\gamma \in 2\pi\mathbb{Z}$ every realization is the reset state and both entropy diagnostics equal 1 bit, while at $2\Omega\gamma \in \pi + 2\pi\mathbb{Z}$ the realizations alternate between two pure states, so $\bar S_A$ is lowered on the half-integer circles of Fig.~\ref{fig:heatmap_Bell} while $S(\bar\rho_A^{(\infty)})$ stays smooth there. For the product-state reset the trajectory accumulates the two phases $Jm$ and $\Omega\gamma m$, and a further collapse occurs on condition (\ref{eq:cond_1}): at $J = \pi k$ the factor $e^{iJm}$ takes at most two values, the two-phase evolution reduces to a single frequency with at most an alternating sign, and $\bar S_A$ is lowered along the horizontal lines $J = \pi k$ of Fig.~\ref{fig:heatmap_00}; $\bar S_A$ is lowered as well along the antiperiodic cross-parity locus $\Omega(\gamma - \alpha) = \pi$, and slightly raised along the antiperiodic loci $\Omega(\gamma + \alpha) \in \pi + 2\pi\mathbb{Z}$. The averaged-state entropy resolves none of these, because condition (\ref{eq:cond_1}) is kinematically suppressed for both reset states and each surviving matrix element averages term by term. On a common locus the two diagnostics also respond differently: at $\Omega\gamma = \pi$ the Bell-reset gap vanishes for every $p$, because the stationary ensemble contains the single state $\ket{\Phi^+}$, while the product-state reset gap remains finite there (0.46 at $p = 0.05$).


\begin{figure*}
\centering
\includegraphics[width=1.0\textwidth]{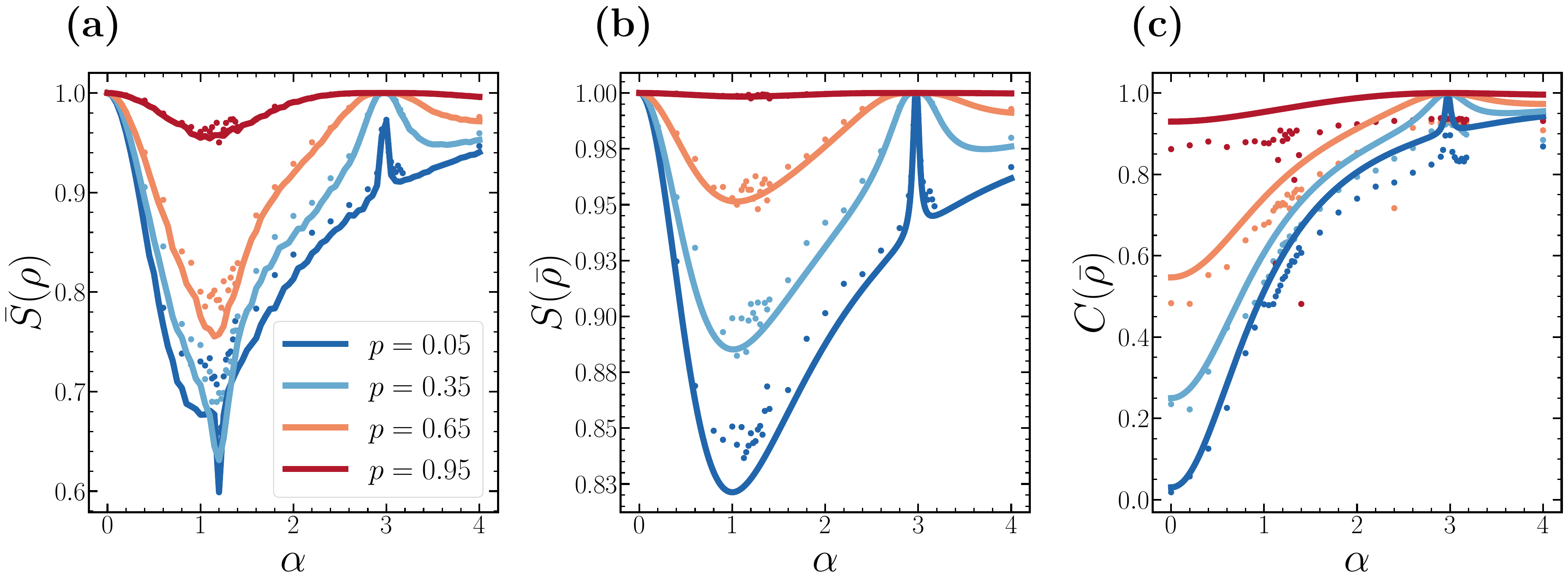}
\caption{Entanglement diagnostics of the Bell-reset stationary state against the \mbox{ibm\_pittsburgh} measurement, from left to right: (a) trajectory-averaged entropy {$\bar S_A$ [Eq.~(\ref{eq:Sbar})], (b) averaged-state entropy $S(\bar\rho_A^{(\infty)})$ [Eq.~(\ref{eq:S_rhobar})], and (c) Wootters concurrence $C(\bar\rho_\infty)$ [Eq.~(\ref{eq:C1})],} as functions of the coupling ratio $\alpha = J/\Omega$ at $\Omega = 1$, for $p \in \{0.05, 0.35, 0.65, 0.95\}$ (from dark blue for $p = 0.05$ to dark red for $p = 0.95$). Solid curves: exact noiseless evaluation of the renewal sum (\ref{eq:NESS}). Symbols: hardware data from qubits 4 and 5 of \mbox{ibm\_pittsburgh}, reconstructed by quantum state tomography with $10^4$ shots per setting, using dynamical decoupling with gate and measurement twirling. The sampling is dense near the resonance window $\alpha \approx 3$. Entropies in bits. At $\alpha = \sqrt{\pi^2 - 1} \approx 2.98$ the stationary state is the pure Bell state and all three observables equal 1. The hardware data at $p = 0.05$ fall short of 1 at this point. In the noiseless protocol every realization at this locus is the Bell state at every age, since $U\ket{\Phi^+} = -\ket{\Phi^+}$ there. The shortfall reflects the hardware noise accumulated over the reset-free intervals, whose mean length $\langle m \rangle = (1-p)/p$ is largest at the smallest reset probability.
}
\label{fig:S_Sbar_DeltaS}
\end{figure*}

\subsection{Comparison with the heatmap data and the hardware experiments}
\label{subsec:heatmap_diagnostics}
\label{subsec:heatmap_match}

{Figures~\ref{fig:heatmap_00} and \ref{fig:heatmap_Bell} show both entropy diagnostics and the Jensen gap across the $(\Omega, J)$ plane at $p = 0.1$, beyond the single coupling ratio and the individual loci examined so far. For the product-state reset, all the entanglement is built by the evolution between resets. At small $\Omega$ the two parity components of $\ket{00}$ are still in phase when the next reset arrives, so both entropies stay small. They rise across a crossover whose midpoint lies where the phase difference of the two components per step, $(\sqrt{J^2 + \Omega^2} - J)\,T$, equals $0.75\,p$, so that they dephase within a typical reset-free interval of order $1/p$ steps. Beyond the crossover the realizations have a median entanglement $\bar S_A$ of 0.46 bits, while the averaged reduced state is close to maximally mixed, and the Jensen gap has a median of 0.49 bits [Fig.~\ref{fig:heatmap_00}(c)]. The Bell reset has no such crossover, because every realization starts maximally entangled. Its averaged-state entropy depends on $\alpha = J/\Omega$ alone, up to corrections of order $p$, and the reduced states of all ages stay close to maximally mixed, so the gap has a median of 0.06 bits [Fig.~\ref{fig:heatmap_Bell}(c)]. The product-state reset thus leaves in $A$ a record of the age more than eight times larger than the Bell reset does.}

On these backgrounds the resonance loci of Sec.~\ref{sec:resonances} appear as narrow arcs and lines. The selection rules of Sec.~\ref{subsec:selection} predict which of these loci appear in $S(\bar\rho_A)$. A feature of $S(\bar\rho_A)$ requires a resonant matrix element of $\bar\rho_\infty$, whereas $\bar S_A$ also shows features wherever the phases accumulated by a realization repeat periodically with its age $m$ (Sec.~\ref{subsec:DeltaS}). Panel (b) of Figs.~\ref{fig:heatmap_00} and \ref{fig:heatmap_Bell} shows the simulated $S(\bar\rho_A)$. For the product-state reset it exhibits dark resonance arcs at radii $\sqrt{J^2 + \Omega^2}$ equal to integer multiples of $\pi$, from condition (\ref{eq:cond_2}), together with lighter arcs, where the entropy is lowered, along the parabolas of condition (\ref{eq:cond_4}), which leave the $J$ axis at $J = \pi k$ and descend with $\Omega$. Condition (\ref{eq:cond_1}) requires the pair $(1, 2)$ and is kinematically suppressed for this reset state, so no resonance lines appear at $J = \pi$, $2\pi$, or $3\pi$. Condition (\ref{eq:cond_3}) requires $\Omega > 2\pi$ for $k = 1$ and lies outside the plotted range. For the Bell reset the same panel shows concentric arcs at the radii $\sqrt{J^2 + \Omega^2} = \pi k$ and no other feature. Only condition (\ref{eq:cond_2}) contributes for the Bell reset, as the selection rule of Sec.~\ref{subsec:selection} states, so this panel is a direct numerical check of the parity structure of Sec.~\ref{sec:resonances}.

Panel (a) of each figure shows $\bar S_A$ for the same parameter ranges, for the two different reset states. The selection rule of Sec.~\ref{subsec:selection} applies to $\bar\rho_\infty$ and therefore to $S(\bar\rho_A)$. The trajectory-averaged entropy $\bar S_A$ is a different observable: it is the average over reset trajectories of the von Neumann entropy of the reduced state of each individual trajectory, not the entropy of the averaged state. Because $\bar S_A$ depends on the pure-state structure of each trajectory and not on the averaged density matrix, it is sensitive to phase-matching conditions that are not visible in $\bar\rho_\infty$. For the product-state reset, panel (a) of Fig.~\ref{fig:heatmap_00} shows dark blue horizontal lines, where $\bar S_A$ is lowered, at $J \approx \pi k$ that are absent from panel (b). These come from condition (\ref{eq:cond_1}), which is kinematically suppressed in $\bar\rho_\infty$ for both reset states because the pair $(1, 2)$ never contributes. The same panel shows the antiperiodic cross-parity curve $\sqrt{J^2 + \Omega^2} - J = \pi$, which follows from the phase collapse described in Sec.~\ref{subsec:DeltaS}. The averaged-state entropy $S(\bar\rho_A)$ is a function of the reduced averaged state alone, so it shows no feature on a locus where no matrix element of $\bar\rho_\infty$ is resonant. The trajectory-averaged entropy $\bar S_A$ is computed from each realization before the average is taken, so the selection rule does not restrict it, and the horizontal lines appear in panel (a). For the Bell reset, panel (a) of Fig.~\ref{fig:heatmap_Bell} adds the half-integer circles. A scan of the closed form (\ref{eq:Sbar_closedform}) along $\Omega = 1$ confirms that the Bell-reset trajectory average has features only at $\gamma \in \tfrac{\pi}{2}\mathbb{Z}$ and none at $J = \pi k$, since it evolves at the single frequency $2\Omega\gamma$. Each feature class is reproduced by evaluating Eqs.~(\ref{eq:S_avgrho}) and (\ref{eq:Sbar_closedform}) at the corresponding loci. The trajectory-averaged entropy therefore resolves condition (\ref{eq:cond_1}), which no quantity computed from $\bar\rho_\infty$ can.

Panel (c) of Figs.~\ref{fig:heatmap_00} and \ref{fig:heatmap_Bell} shows the Jensen gap itself, by Eq.~(\ref{eq:holevo}) the most a measurement on subsystem $A$ can learn about the age. Where $\bar S_A$ is lowered and $S(\bar\rho_A)$ is not, the reduced states of different ages differ more, and the gap is raised. This happens on the lines $J = \pi k$ and the curve $\sqrt{J^2 + \Omega^2} - J = \pi$ of Fig.~\ref{fig:heatmap_00} and on the half-integer circles of Fig.~\ref{fig:heatmap_Bell}. Along the parabolas of condition (\ref{eq:cond_4}), the gap is lowered. On the circles $\sqrt{J^2 + \Omega^2} = \pi k$ of the Bell reset, every realization is the Bell state at every age, so the gap vanishes and subsystem $A$ holds no information about the age there.

The hardware measurement follows one line of the plane of Figs.~\ref{fig:heatmap_00} and \ref{fig:heatmap_Bell}, the line $\Omega = 1$. This line crosses the circle $\sqrt{J^2 + \Omega^2} = \pi$ of condition (\ref{eq:cond_2}) at $\alpha = \sqrt{\pi^2 - 1} \approx 2.98$, inside the resonance window of Sec.~\ref{sec:resonances}. Fig.~\ref{fig:S_Sbar_DeltaS} compares the two entropy diagnostics and the concurrence with the measurement on \mbox{ibm\_pittsburgh}, for the Bell reset at $p \in \{0.05, 0.35, 0.65, 0.95\}$ across $\alpha \in [0, 4]$ at $\Omega = 1$. The solid curves are the exact noiseless evaluation of the renewal sum (\ref{eq:NESS}). The symbols are the hardware data, reconstructed by quantum state tomography from the protocol of Sec.~\ref{sec:model}, ensemble-averaged for panels (b) and (c) and trajectory-resolved for panel (a). Both entropy diagnostics follow the noiseless curves closely across the measured range, with small residual deviations and no fitted parameter. The concurrence data lie below the noiseless prediction across most of that range. The exception is the $p = 0.05$ and $p = 0.35$ data near $\alpha = 0$, where the noiseless concurrence approaches zero and the data lie slightly above it. At $\alpha = \sqrt{\pi^2 - 1}$, where $\Omega\gamma = \pi$ and the stationary state is the pure Bell state, both entropy diagnostics equal 1 bit, the concurrence equals 1, and the Jensen gap, the difference between panels (b) and (a), vanishes; the hardware sampling is dense in this window. 
The asymmetry between the entropy diagnostics and the concurrence is quantified in Sec.~\ref{sec:noise}.

\subsection{Significance for experimental diagnostics on noisy hardware}
\label{subsec:experimental_significance}

The two entropy diagnostics differ in measurement cost. $S(\bar\rho_A^{(\infty)})$ follows from ensemble-averaged tomography, with shot counts set by the precision required on one reconstructed density matrix. $\bar S_A$ requires trajectory-resolved reconstruction. Shot counts are split across the distinct reset histories in the stationary ensemble, so each history class ($m$ bins in Eq.~\eqref{eq:Sbar_closedform}) is reconstructed with lower statistical precision than the single averaged state. At the precision of Fig.~\ref{fig:S_Sbar_DeltaS}, both entropy diagnostics follow the noiseless theory. As discussed in Sec.~\ref{sec:introduction}, no shot is discarded in this splitting. Since the reset times are drawn and recorded classically, the binning by reset history is a partition of the data rather than a postselection, in contrast to monitored circuits, where trajectory-resolved quantities require postselection on a measurement record whose probability falls exponentially with the number of measurements~\cite{2023_Koh_MIPT}.

The diagnostics of the averaged state also respond differently to hardware noise. The first-order noise response of the averaged-state entropy is suppressed relative to that of the concurrence, because the partial trace removes the Pauli-weight sectors that capture the dominant response. Section~\ref{sec:noise} derives this partial-trace projection mechanism and quantifies both responses, including the fitted effective depolarizing strength; it is the reason panel (b) of Fig.~\ref{fig:S_Sbar_DeltaS} follows the noiseless theory while panel (c) is displaced. The agreement of the trajectory-averaged entropy in panel (a) is an observation at the achieved precision.

\section{Exact calculations for the transverse-field Ising chain}
\label{sec:exact_chain}

The two-qubit analysis of Secs.~\ref{sec:model} through \ref{sec:diagnostics} is tractable in closed form, but it does not answer whether the diagnostic distinction identified there persists in the many-body context. The Jensen gap of Eq.~(\ref{eq:DeltaS_def}) can be generalized to any subsystem $A$ of a chain, $\Delta S_A = S(\bar\rho_A) - \bar S_A \geq 0$, and what the two-qubit results do not resolve is how it scales with system size: whether it stays a bounded local quantity, or becomes extensive as the subsystem grows. We answer this for the transverse-field Ising chain by evaluating the reset stationary state exactly for chains of up to $N = 12$ sites. The resonance classification of Sec.~\ref{sec:resonances} admits a chain analog. Because the chain is integrable, the many-body resonances are integer combinations of the free-fermion single-particle energies, and the number of resonance loci grows with system size, as sketched in Appendix~\ref{app:chain_phase_matching}. The treatment in the main text is restricted to the Jensen gap, which does not depend on this resonance structure.

\subsection{Chain Hamiltonian and exact evaluation}
\label{subsec:chain_exact}

The extension of the two-qubit Hamiltonian (\ref{eq:hamiltonian_2qubit}) to a chain of $N$ sites with open boundary conditions and the same couplings is
\begin{equation}
H_N = -J \sum_{i=1}^{N-1} \sigma_i^z \sigma_{i+1}^z
      + \frac{\Omega}{2} \sum_{i=1}^{N} \sigma_i^x,
\label{eq:chain_hamiltonian}
\end{equation}
the standard transverse-field Ising chain. The reset protocol of Sec.~\ref{sec:model} is unchanged, with $U_N = e^{-iH_NT}$ and $T = 1$, and we retain the two reset states in their chain form: the product-state reset $\ket{\psi_0} = \ket{0}^{\otimes N}$, and the Bell reset taken as the tiled product of nearest-neighbor Bell pairs, $\ket{\psi_0} = \ket{\Phi^+}_{1,2} \otimes \ket{\Phi^+}_{3,4} \otimes \cdots$, which reduces to the single Bell pair at $N = 2$, and is defined for even $N$, which is the case throughout the paper. The subsystems are taken at the center of the chain to limit boundary effects: the central site for the local gap, and the half chain to one side of the central bond for the half-chain gap.

The closed forms of Sec.~\ref{subsec:rho_construction} do not extend to $N > 2$, but none of the diagnostics requires the full stationary state. By linearity of the partial trace, the reduced stationary state $\bar\rho_A = \mathrm{Tr}_{\bar A}\,\bar\rho_\infty$ inherits the renewal structure of (\ref{eq:NESS}), with $\rho_A(m) = \mathrm{Tr}_{\bar A}[U_N^m \rho_0 U_N^{\dagger m}]$ the reduced state of one exactly evolved pure state. The operator $\bar\rho_\infty$, of dimension $2^N \times 2^N$, is never constructed: a state vector of dimension $2^N$ is propagated exactly, reduced to $A$, and accumulated with the geometric weights. The one family $\{\rho_A(m)\}$ then gives both entropy diagnostics, (\ref{eq:S_rhobar}) and (\ref{eq:Sbar}), hence the Jensen gap. The geometric sum is truncated as in Sec.~\ref{subsec:Sbar_closedform}.

The evaluation is exact to machine precision, and its reach is set by the growth of the state-vector dimension with $N$. We report chains up to $N = 12$, for which the state vector has dimension $4096$ and the central-bond reduced state is $64 \times 64$; larger chains are deferred to follow-up work. At $N = 2$ the chain evaluation reproduces the closed-form reduced state (\ref{eq:rhoA_inf}) and the trajectory-averaged entropy (\ref{eq:Sbar_closedform}) to machine precision. Direct iteration of the update (\ref{eq:update}) on the chain reaches the same stationary state, with a deviation that decays as $(1-p)^n$ in the number of steps $n$, as the two-qubit populations do in Fig.~\ref{fig.1:circuit}(c).

\subsection{The Jensen gap across the coupling range}
\label{subsec:chain_diagnostics}

\begin{figure}
\centering
\includegraphics[width=1.0\linewidth]{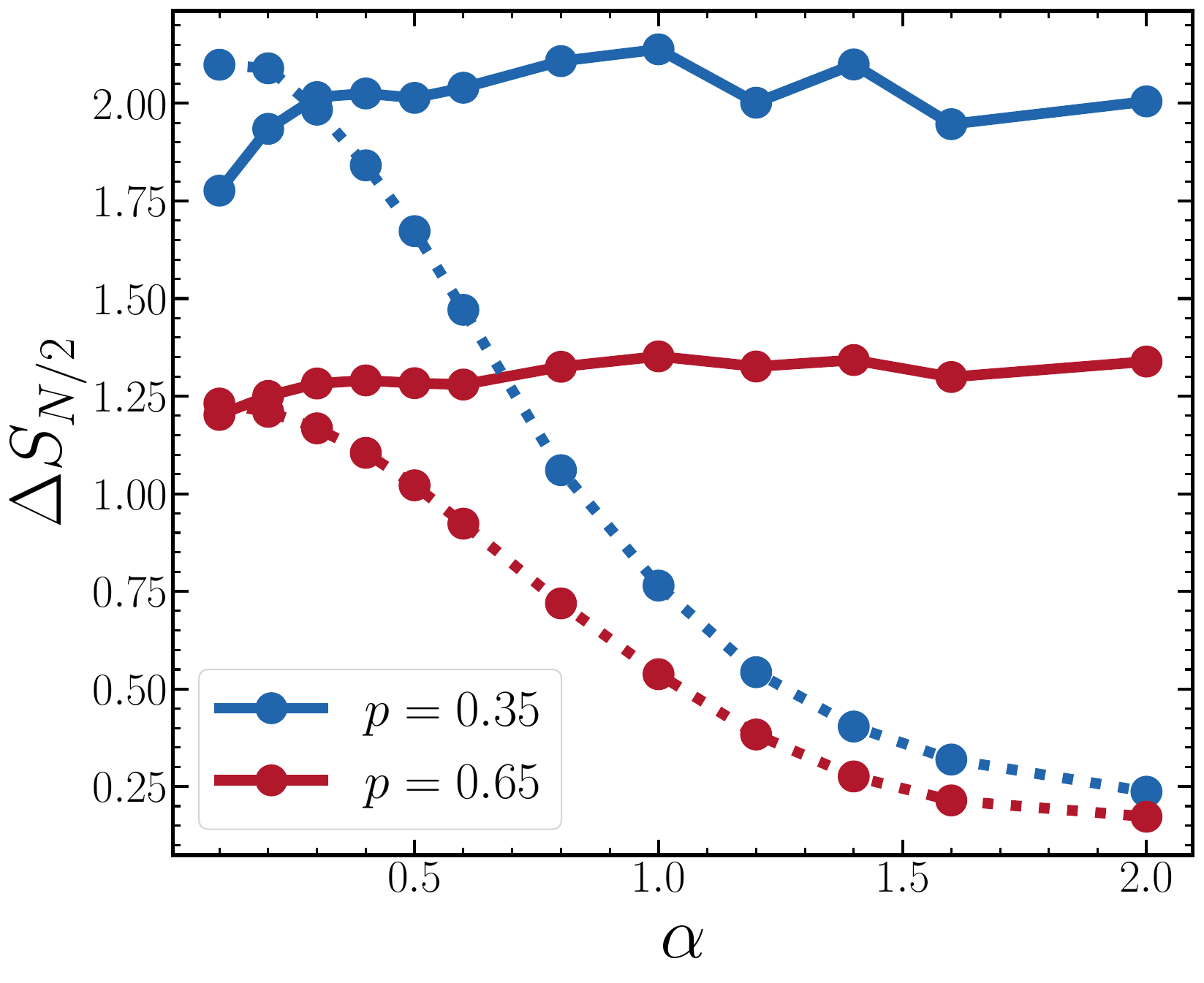}
\caption{Half-chain Jensen gap {[Eq.~(\ref{eq:DeltaS_def})]} of the reset stationary state at $N = 12$ versus the coupling ratio $\alpha = J/\Omega$, for the tiled-Bell reset (solid) and the product-state reset (dotted), at $p = 0.35$ and $0.65$ (dark blue and red). Markers are the exact evaluation; entropies in bits. The two coupling ratios of Fig.~\ref{fig:chain_scaling} are read off this profile: comparable reset states at $\alpha = 0.2$, an order-of-magnitude separation at $\alpha = 2.0$.
}
\label{fig:chain_gap_alpha}
\end{figure}

Figure~\ref{fig:chain_gap_alpha} shows the half-chain Jensen gap at $N = 12$ as a function of the coupling ratio, for both reset states at $p \in \{0.35, 0.65\}$. The gap of Sec.~\ref{sec:diagnostics} is present at chain scale across the whole range of $\alpha$ shown, and the two reset states give profiles that match their roles at two qubits. For the tiled-Bell reset the gap stays between $1.2$ and $2.14$, with a broad maximum near $\alpha = 1$ where the Ising and transverse-field scales are comparable. For the product-state reset the gap falls monotonically with $\alpha$, from $2.1$ at $\alpha = 0.1$ to $0.24$ at $\alpha = 2.0$ for $p = 0.35$: the unitary builds correlations from an unentangled product state, and does so most effectively where the transverse field dominates, while the Bell reset supplies entanglement that the dynamics redistributes at every coupling. The two coupling ratios of the size-scaling analysis below are read off this profile: at $\alpha = 0.2$ the two reset states are comparable, and at $\alpha = 2.0$ they are separated by an order of magnitude.

\subsection{Size scaling of the Jensen gap}
\label{subsec:chain_scaling}

\begin{figure}
\centering
\includegraphics[width=1.0\linewidth]{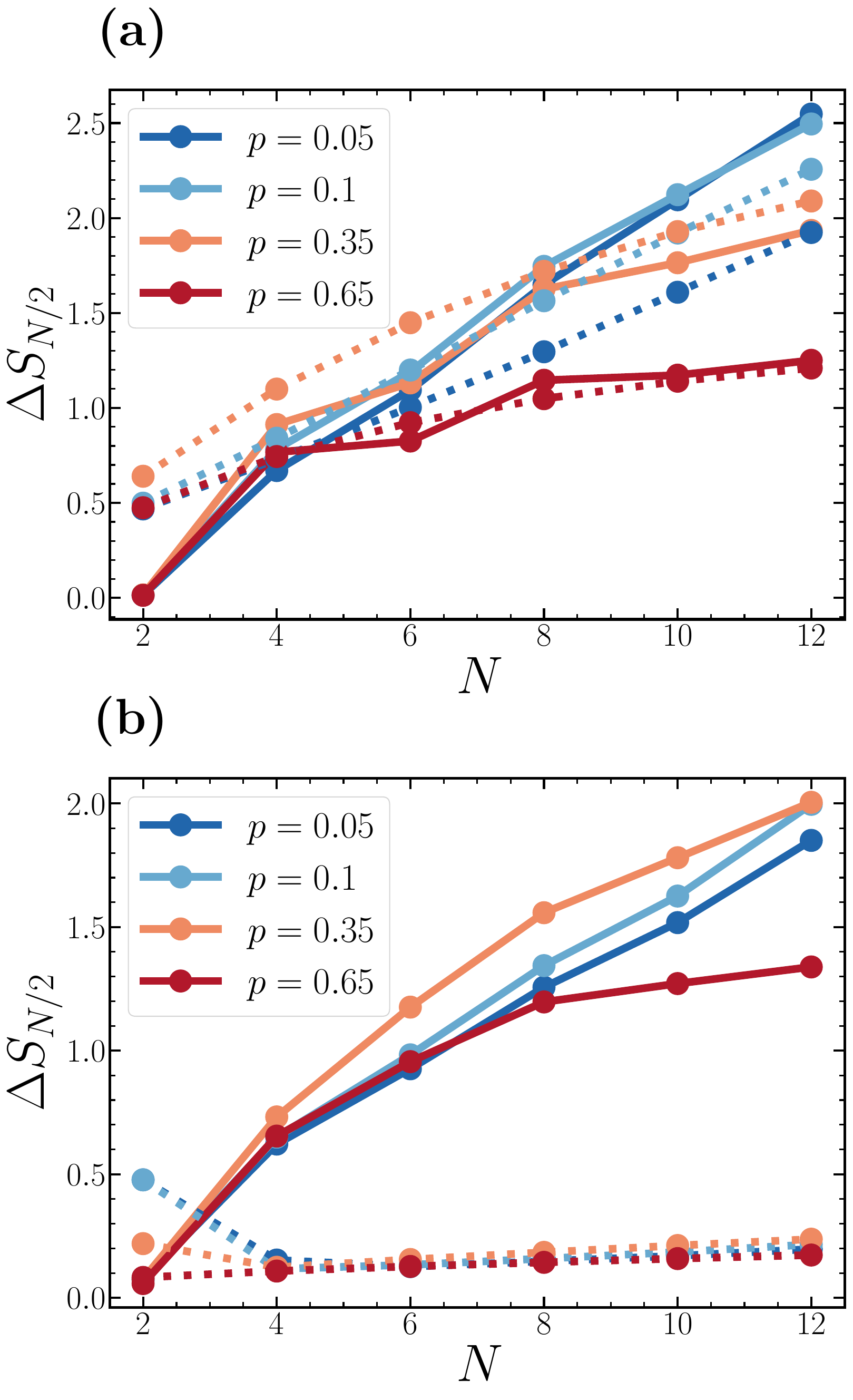}
\caption{Size scaling of the half-chain Jensen gap [Eq.~(\ref{eq:DeltaS_def})]. $\Delta S_{N/2}$ versus chain length $N$ for the Bell reset (solid) and the product-state reset (dotted), at (a) $\alpha = 0.2$ and (b) $\alpha = 2.0$, for $p = 0.05$, $0.1$, $0.35$, and $0.65$ (from dark blue for $p = 0.05$ to red for $p = 0.65$). Markers are the exact evaluation; entropies in bits. The gap grows monotonically with $N$ for the Bell reset and for the product-state reset at $\alpha = 0.2$. At $\alpha = 2.0$ the product-state reset gap stays below $0.24$ for $N \geq 4$, so the magnitude at chain scale is selected by the reset state.
}
\label{fig:chain_scaling}
\end{figure}

Figure~\ref{fig:chain_scaling} shows the half-chain Jensen gap $\Delta S_{N/2}$ as a function of chain length for both reset states at matched parameters: the same reset probabilities $p \in \{0.05, 0.1, 0.35, 0.65\}$ at $\alpha = 0.2$ [Fig.~\ref{fig:chain_scaling}(a)] and $\alpha = 2.0$ [Fig.~\ref{fig:chain_scaling}(b)]. The gap grows monotonically with $N$ for the Bell reset at both coupling ratios and for the product-state reset at $\alpha = 0.2$; at the smaller reset probabilities the increments show no sign of saturation within the chains the exact method reaches, while at $p = 0.65$ the growth slows above $N = 8$. For the product-state reset at $\alpha = 2.0$ the gap drops from its two-qubit value at $N = 4$ for $p \leq 0.35$, rises from it for $p = 0.65$, and then grows slowly with $N$ while staying below $0.24$ for $N \geq 4$. We do not fit a scaling form, because the growth is capped by the bound $h(p)/p$ of Eq.~(\ref{eq:age_entropy_bound}), and resolving how it approaches that cap requires more sites than the exact method reaches here; this is deferred to follow-up work.

The magnitude of the growth is selected by the reset state in the Ising-dominated regime. At $\alpha = 2.0$ the Bell-reset gap reaches $2.0$ at $N = 12$ for $p = 0.35$ while the product-state reset gap stays below $0.24$ for $N \geq 4$, an order-of-magnitude separation at matched parameters. At $\alpha = 0.2$ the two reset states grow comparably, and the largest gap in the data set, $2.55$ at $N = 12$, occurs for the Bell reset at the smallest reset probability. At $N = 12$ the gap is largest at an intermediate reset probability in three of the four cases, at $p = 0.1$ or $0.35$, as the two-qubit gap is in Sec.~\ref{subsec:DeltaS}; for the Bell reset at $\alpha = 0.2$ it decreases with $p$ over the values shown. The corresponding single-site gap $\Delta S_1$ (not shown) reaches its largest value for $N \geq 4$, $0.556$, for the product-state reset at $\alpha = 0.2$, $p = 0.35$, where it is already close to converged by $N = 4$. Elsewhere in the data set it typically continues to decrease out to $N = 12$, most markedly at $\alpha = 2.0$, and stays below $0.011$ for the Bell reset at either coupling ratio for $N \geq 6$ (0.015 at $N = 4$, $\alpha = 2.0$, and $p = 0.05$).

The contrast between the bounded local gap and the half-chain gap that grows with $N$ has a direct information-theoretic origin. By Eq.~(\ref{eq:holevo}), the Jensen gap of a subsystem $A$ bounds the information about the age that any measurement on $A$ alone can extract. It is bounded by the entropy of the averaged state, $\Delta S_A \leq S(\bar\rho_A) \leq \ell$ for a subsystem of $\ell$ sites (in bits), so the single-site gap can never exceed $1$ and its saturation is forced by dimension alone. The gap is also bounded by the Shannon entropy of the age distribution, because the entropy of a mixture never exceeds the mean entropy of its components plus the entropy of their weights~\cite{2010_Nielsen_qcqi,1978_Wehrl_entropy}. For the age weights $w_m$ of Eq.~(\ref{eq:age_law}), $\log_2 w_m = \log_2 p + m \log_2(1-p)$, and averaging with the mean age $\langle m \rangle = (1-p)/p$ gives
\begin{align}
\Delta S_A &\leq -\sum_{m=0}^{\infty} w_m \log_2 w_m \nonumber \\
&= -\log_2 p - \frac{1-p}{p}\,\log_2(1-p) = \frac{h(p)}{p},
\label{eq:age_entropy_bound}
\end{align}
in bits, with $h$ the binary entropy function of Eq.~(\ref{eq:S_avgrho}). This bound does not depend on $N$, so $\Delta S_{N/2}$ cannot keep growing in proportion to $N$. At $p = 0.35$ the bound is $2.67$ bits, of which the Bell-reset half chain reaches $2.0$ bits at $N = 12$ and $\alpha = 2.0$. At $p = 0.65$ the bound is $1.44$ bits, close to the Bell-reset values of Fig.~\ref{fig:chain_scaling} at $N = 12$ (1.25 and 1.34 bits), which is consistent with the slower growth at that reset probability. The monotonic growth of $\Delta S_{N/2}$, in contrast, states that, over the sizes studied, the half chain stores a record of the reset age that grows with $N$. Between resets the unitary spreads correlations across the central bond, so reduced states of different ages remain mutually distinguishable there as $N$ grows. The single-site gap, which can never exceed 1 bit, decreases with $N$ in most of the data set. No quantity computed from $\bar\rho_\infty$ retains this record, since the age is averaged out of the stationary state by construction. The same noncommutativity separates the trajectory-averaged entropy from the averaged-state entropy in the measurement-induced transition literature \cite{2019_Skinner_MIPT,2019_Li_hybrid,2020_Gullans_purification}, where the trajectory-resolved entanglement holds the extensive information of the volume-law phase.

\section{Noise sensitivity of the two diagnostics}
\label{sec:noise}

The diagnostics of Secs.~\ref{sec:diagnostics} and \ref{sec:exact_chain} were evaluated for the ideal protocol, with the state evolved exactly and no decoherence. On real hardware, noise changes the states produced by the reset protocol and can affect the measures differently. Across the measured range, both entropies remain close to the noiseless predictions with no fitted parameter, whereas concurrence lies below its noiseless prediction across most of that range (Fig.~\ref{fig:S_Sbar_DeltaS}). Understanding this difference requires examining how noise changes both the reduced state of one qubit, which determines the averaged-state entropy, and the two-qubit correlations that also enter concurrence. We investigate these changes using a depolarizing model that includes the unitary evolution between noise applications. The calculation compares the noise responses of averaged-state entropy and concurrence.

\subsection{Noise model}
\label{subsec:noise_model}

We model the aggregate effect of hardware imperfections by a product of identical single-qubit depolarizing channels,
\begin{equation}
\mathcal{E}_\epsilon = \mathcal{D}_\epsilon \otimes \mathcal{D}_\epsilon,
\qquad
\mathcal{D}_\epsilon[\rho] = (1 - \epsilon)\,\rho
   + \epsilon\,\frac{\mathbb{I}}{2},
\label{eq:depolarizing}
\end{equation}
where $\epsilon \in [0, 1]$ is the depolarizing strength. In the Pauli basis $\mathcal{D}_\epsilon$ fixes the identity and multiplies each single-qubit Pauli by $1 - \epsilon$, so a two-qubit Pauli string of weight $w$, the number of non-identity factors, is damped by $(1 - \epsilon)^w$ per application of the channel. Applied after each unitary step, the channel defines the noisy stationary state through the fixed-point equation
\begin{equation}
\bar\rho_\infty^{(\epsilon)} = (1 - p)\,
  \mathcal{E}_\epsilon\!\left[U\bar\rho_\infty^{(\epsilon)}
  U^\dagger\right] + p\,\rho_0,
\label{eq:noisy_fixedpoint}
\end{equation}
which reduces to the noiseless fixed point (\ref{eq:fixedpoint}) at $\epsilon = 0$. This is a deliberately minimal description: one parameter, no distinction between gate, idling, and readout errors, and no relaxation or dephasing asymmetry. For monitored recurrence on the same family of processors, Ma \textit{et al.} \cite{2026_Ma_noisy} find that a depolarizing channel does not reproduce the data near revivals, whereas generalized amplitude damping with asymmetric per-cycle flip probabilities $p_{0\to1}\approx10^{-3}$ and $p_{1\to0}\approx10^{-2}$ does. The depolarizing model describes the present data well, though not perfectly. The related hardware study of Murauer~\textit{et al.}~\cite{2026_Murauer_NESS} finds that a symmetry-based reset-state-error model describes its conditional-reset data better than depolarizing, phase-damping, amplitude-damping, and correlated-$ZZ$ channels; a device-calibrated treatment along those lines is left to future work.

\subsection{Partial-trace origin of the asymmetry}
\label{subsec:structural_origin}

The partial trace is a projection in the Pauli basis. It satisfies $\mathrm{Tr}_B[\sigma_a \otimes \sigma_b] = \mathrm{Tr}[\sigma_b]\,\sigma_a$, which vanishes unless $b = \mathbb{I}$, so the reduced state $\bar\rho_A$ depends on four of the sixteen Pauli coefficients of the two-qubit state: the identity and the three weight-one strings supported on qubit 1. The concurrence, evaluated on the full two-qubit state, depends on all sixteen, including the nine weight-two coefficients. The channel, in addition, damps by weight: under one application of (\ref{eq:depolarizing}), weight-one coefficients are multiplied by $1 - \epsilon$ and weight-two coefficients by $(1 - \epsilon)^2$, so to first order in $\epsilon$ the coefficients that survive the partial trace are damped at rate one, while the weight-two coefficients, which only the concurrence reads, are damped at rate two.

These two statements concern a single application of the channel. In the stationary state the channel acts once per time step and the unitary evolution between applications mixes the Pauli-weight sectors, so the per-step grading does not transfer unchanged to $\bar\rho_\infty^{(\epsilon)}$; how much of it survives is a quantitative question. We answer it numerically from the exact solution of (\ref{eq:noisy_fixedpoint}), with the concurrence derivative taken as $\epsilon \to 0^+$ because $C$ is not differentiable where a Wootters eigenvalue vanishes: across the Bell-reset scan of Fig.~\ref{fig:S_Sbar_DeltaS} and the product-state reset at $p = 0.1$ for $0 < \alpha \leq 4$, the first-order responses satisfy $|\partial S/\partial\epsilon| \leq 0.33\,|\partial C/\partial\epsilon|$, with $|\partial S/\partial\epsilon| = 0.95$ bits and $|\partial C/\partial\epsilon| = 12.7$ at the operating point of Fig.~\ref{fig.1:circuit}. The suppression is a difference of rates, with both responses finite. The first-order correction and the slopes at the operating points are collected in Appendix~\ref{app:noise_perturbation}.

In the same calculation at $p = 0.1$, the entropy slope of the product-state reset and the concurrence slope of the Bell reset are largest on the circular locus (\ref{eq:cond_2}), and the entropy slope of the Bell reset vanishes there. For the product-state reset at $p = 0.1$, $|\partial S/\partial\epsilon|$ rises toward the resonance window, from $0.28$ bits at $\alpha = 2.5$ to $1.46$ bits at the circular locus (\ref{eq:cond_2}), $\alpha = \sqrt{\pi^2 - 1}$, since the stationary coherences that the channel damps are largest there. For the Bell reset the behavior at the same locus follows from an exact property of the protocol: at $\Omega\gamma = \pi$ the step unitary acts on the reset state as $U\ket{\Phi^+} = -\ket{\Phi^+}$, the stationary state is $\ket{\Phi^+}\!\bra{\Phi^+}$ itself, and the reduced state is maximally mixed, so the first-order entropy response vanishes, while $|\partial C/\partial\epsilon|$ is largest there (a slope of $27$ at $p = 0.1$). The two diagnostics are thus predicted to disagree most strongly on the resonance loci. The present data sample the resonance window densely; a quantitative test of this prediction is left to future work.

\subsection{Comparison with experiment}
\label{subsec:experimental_confirmation}

The measured entropy diagnostics follow the noiseless closed forms at all four reset probabilities with no fitted parameter, and they reproduce the resonance structure near $\alpha = \sqrt{\pi^2 - 1}$ (Fig.~\ref{fig:S_Sbar_DeltaS}, Sec.~\ref{subsec:heatmap_diagnostics}). Given the noise levels of current hardware, this agreement is the first half of the observed asymmetry.

\begin{figure}
\centering
\includegraphics[width=1.0\linewidth]{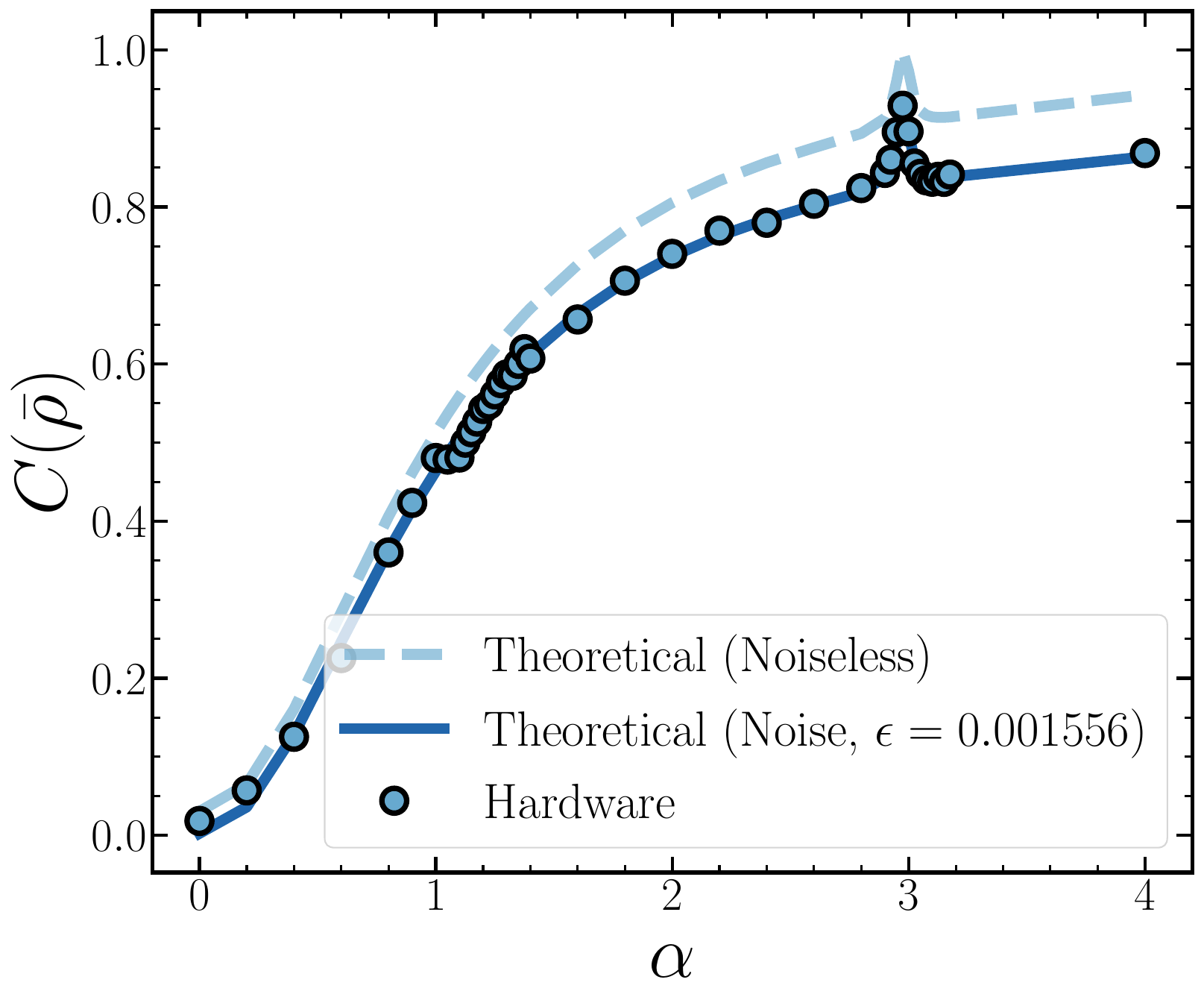}
\caption{Stationary-state concurrence {$C(\bar\rho_\infty)$ [Eq.~(\ref{eq:C1})]} of the Bell reset at $p = 0.05$ on ibm\_pittsburgh: noiseless theory (dashed), the depolarizing model at the fitted strength $\epsilon = 1.56\times10^{-3}$ per step (solid line), and the hardware data (markers). The fit (COBYQA, bounds $[0,1]$) gives a mean-square error of $9.8\times10^{-5}$ against the $p=0.05$ data.
}
\label{fig:concurrence_fit}
\end{figure}

The second half is the concurrence. Figure~\ref{fig:concurrence_fit} shows the measured stationary-state concurrence at $p = 0.05$ against the noiseless theory and against the depolarizing model with the noise strength as the only fitted parameter. A fit minimizing the mean-square error (COBYQA \cite{2022_Ragonneau_dfo, 2025_Ragonneau_cobyqa}, bounds $[0, 1]$, random initialization, trust-region convergence) gives $\epsilon = 1.56\times 10^{-3}$ per step, with a mean-square error of $9.8\times 10^{-5}$, against $4.1\times 10^{-3}$ for the noiseless theory. The single-parameter model captures the displacement of the data from the noiseless curve across the whole measured range, $\alpha$ from 0 to 4, with a root-mean-square deviation of 0.010. It also reproduces the shortfall below 1 at the circular locus, where it calculates 0.916 against the measured 0.929. We do not attach an uncertainty to $\epsilon$ here, and the quoted value is the fit to the $p = 0.05$ data set.

\section{Summary and discussion}
\label{sec:discussion}

For the transverse-field Ising model under discrete-time stochastic resetting, we compared the averaged-state entropy $S(\bar\rho_A^{(\infty)})$, the trajectory-averaged entropy $\bar S_A$, and the Wootters concurrence $C(\bar\rho_\infty)$ of the stationary state, with exact two-qubit calculations, measurements on an IBM superconducting processor, and exact evaluations for chains of up to $N = 12$ sites. For two qubits, a coherence between energy eigenstates $i$ and $j$ survives the averaging unchanged when $(E_i - E_j)T \in 2\pi\mathbb{Z}$, provided the reset state overlaps with both eigenstates; the energy spectrum determines where this phase matching occurs, while reset-state overlaps and parity determine which pairs contribute (Secs.~\ref{sec:resonances} and \ref{subsec:selection}). The trajectory-averaged entropy also shows features where the evolution follows a repeated pattern, such as alternation between two pure states, with no corresponding resonance in the averaged state, because the entropy is computed before the average (Sec.~\ref{subsec:DeltaS}).

The Jensen gap $\Delta S$ of Eq.~(\ref{eq:DeltaS_def}) is the extra uncertainty about subsystem $A$ that comes from not knowing the reset age, and equals the Holevo information of Eq.~(\ref{eq:holevo}). It bounds what a measurement of subsystem $A$ can reveal about the number of steps since the last reset, and it vanishes exactly when $A$ has the same reduced state at every age. It does not measure the entanglement lost by averaging (Secs.~\ref{subsec:two_protocols} and \ref{sec:exact_chain}). For a single site the gap cannot exceed one bit; for half of the chain it increases with $N$ over the sizes studied (Fig.~\ref{fig:chain_scaling}), for the tiled-Bell reset at both $\alpha = 0.2$ and $\alpha = 2.0$ and for the product-state reset at $\alpha = 0.2$, and it is bounded for all $N$ by $h(p)/p$, the Shannon entropy of the reset-age distribution (Sec.~\ref{subsec:chain_scaling}).

On \mbox{ibm\_pittsburgh}, both entropies closely follow the noiseless predictions across the Bell-reset measurements with no fitted parameter, whereas the concurrence lies below the noiseless result across most of the measured range. A product of identical single-qubit depolarizing channels with one fitted strength captures the overall concurrence suppression at $p = 0.05$. The difference arises because the partial trace retains only the identity and the single-qubit Pauli terms on $A$, which the channel damps more weakly than the two-qubit correlations the concurrence also reads (Secs.~\ref{subsec:structural_origin} and \ref{subsec:experimental_confirmation}).

Because the reset record is classical, $\bar S_A$ is obtained from repeated measurements at each age with the known geometric weights, without postselecting shots that share a complete measurement record. Resolving the ages $m$, however, divides the measurement statistics among one reconstructed state per age and increases the sampling cost (Sec.~\ref{subsec:experimental_significance}). The interpretation of $\bar S_A$ as mean entanglement also requires pure realizations; with noise, a reduced-state entropy can include mixedness that is unrelated to entanglement.

The order of entropy evaluation and ensemble averaging also matters in monitored circuits, where a measurement-induced transition in trajectory entanglement need not appear in the entropy of the state averaged over all measurement outcomes \cite{2019_Skinner_MIPT,2019_Li_hybrid,2020_Gullans_purification,2026_Hamazaki_review}. Global resets and measurements conditioned on their outcomes generate different ensembles. The connection concerns the distinction between averaging states and averaging their entropies; our results do not establish a measurement-induced transition. Hardware studies of monitored dynamics \cite{2023_Koh_MIPT} and postselection-free probes \cite{2025_Kamakari_XEB} examine how trajectory information can be accessed experimentally. In the ideal reset protocol, this task is simplified because a single age specifies the state. Random projective measurements have also been studied through their effect on first-detection statistics \cite{2023_Kulkarni_detection}. Whether the transitions found in monitored systems survive decoherence is a related question \cite{2025_Qian_protect}, although our two-qubit noise comparison does not test the stability of a many-body transition.

The phase-matching loci of Sec.~\ref{sec:resonances} also govern a different protocol on the same two qubits. Under stroboscopic projective monitoring of a target state, the mean number of measurements to the first detected return is an integer equal to the number of distinct phases $e^{-iE_kT}$ among the eigenstates the target overlaps \cite{2013_Grunbaum_recurrence,2017_Friedman_walks}. It drops by one wherever two of those phases coincide \cite{2025_Yin_resonances,2026_Liu_recurrence}. The condition is that of Eq.~(\ref{eq:resonance_intro}). A pair $(E_i,E_j)$ lowers the return time from $\ket{\psi_0}$ exactly when $(E_i-E_j)T\in2\pi\mathbb{Z}$ and $\braket{E_i|\psi_0}\braket{\psi_0|E_j}\neq0$, which is the condition under which the coherence $\bra{E_i}\bar\rho_\infty\ket{E_j}$ of Eq.~(\ref{eq:eigenbasis}) survives the averaging in full. For the product-state reset the return time from $\ket{00}$ equals $3$ away from the loci and $2$ on each of the three loci its parity admits; for the Bell reset it equals $2$ and drops to $1$ on the circular locus, where $U\ket{\Phi^+}=-\ket{\Phi^+}$. The surviving coherences of the reset stationary state and the monitored return time are thus determined by one spectral condition, and both are read from the measurement record without postselection. On noisy hardware the two respond differently to noise. Near revivals the integer return time is strongly renormalized by asymmetric relaxation \cite{2026_Ma_noisy}, whereas the resonance positions of the stationary state are fixed by the spectrum and only their visibility is damped (Sec.~\ref{sec:noise}).

The depolarizing model omits differences between relaxation and dephasing rates, correlated gate errors, and errors in resetting and mid-circuit measurements \cite{2025_Koh_readout,2026_Murauer_NESS}. A model calibrated to the device would test how these processes affect the observed contrast. Partial tracing retains the same Pauli terms for other noise channels, but those channels need not damp two-qubit correlations more strongly than single-qubit terms. Thus, the relative noise sensitivities found here cannot be assumed for other noise models.

Looking forward, beyond two qubits, classifying which many-body resonances contribute remains an open problem. The free-fermion description of the Ising chain expresses the energy differences of Eq.~(\ref{eq:resonance_intro}) as combinations of single-particle energies (Appendix~\ref{app:chain_phase_matching}), while reset-state overlaps determine which coherences are present. In a future extension of this work, adding a longitudinal field could test how the resonance features in the two entropies and concurrence change when integrability is broken. The spectral phase-matching condition remains applicable. Resetting into a scarred subspace, in which the system can retain memory of its initial state, provides another nonintegrable setting \cite{2026_Gotta_scars}. Entanglement negativity, a measure of entanglement for mixed states, would extend the comparison to larger subsystems. For a specified sensing task, quantum Fisher information could assess the attainable estimation precision and test whether the reset parameters that favor entanglement also favor sensing. Such a comparison would need to account for preparation time and access to the reset record. Choosing reset intervals from a nongeometric distribution would change the age weights. If the mean interval diverges, the age distribution can remain dependent on the observation time, so a stationary age ensemble cannot be assumed. Determining the long-time ensemble would be necessary before extending the entropy comparison to this regime and exploring how rare, long intervals between resets affect time averages \cite{2023_Barkai_ergodic,2023_Afek_anomalous}.

\begin{acknowledgments}
This work benefitted from discussions at the ICTP School on Quantum Many-Body Phenomena out of Equilibrium: from Chaos to Criticality (smr 3867) and the concurrent ICTP Workshop on Dynamics of Monitored Quantum Many-Body Systems (smr 3868), held in Trieste in August 2023 and continued at the International Workshop Quantum Meets Classical, held at the Max Planck Institute for the Physics of Complex Systems in Dresden in October 2024.
This research was supported in part by the International Centre for Theoretical Sciences (ICTS) for the program Quantum Trajectories (code: ICTS/qt2025/01).
M.K. acknowledges support from the Department of Atomic Energy, Government of India, under Project No. RTI4001.
M.S.R. acknowledges support from the Max Planck Institute of Quantum Optics (MPQ) and the Munich Center for Quantum Science and Technology (MCQST), and thanks the Research Institute CODE, University of the Bundeswehr Munich, for hosting her in Munich during her visits.
This research was supported by the Israel Science Foundation (grant No. 2311/25) (EB).
We acknowledge the use of IBM Quantum services for this work. The views expressed are those of the authors and do not reflect the official policy or position of IBM or the IBM Quantum team.
\end{acknowledgments}

\section*{Data availability}
The code for constructing the reset stationary states and evaluating the two entropies and concurrence is available in Ref.~\cite{2026_SinghRoy_code}.

\appendix

\section{The $p \to 0$ limit of the Jensen gap}
\label{app:jensen_limit}

The $p \to 0$ limit of the Jensen gap quoted in Sec.~\ref{subsec:DeltaS} rests on one summation fact: the renewal average $p \sum_{m=0}^{\infty} (1-p)^m f(m)$ is the Abel mean of the sequence $f(m)$, and for a bounded sequence whose time average $\lim_{M\to\infty} M^{-1} \sum_{m=0}^{M-1} f(m)$ exists, the Abel mean converges to that time average as $p \to 0$ \cite{1949_Hardy_divergent}. Both entropy diagnostics are built from such means: the averaged-state entropy from the Abel mean of the matrix elements of $\rho_A(m)$, with the entropy functional applied afterwards by continuity, and the trajectory-averaged entropy from the Abel mean of the bounded sequence of single-realization entropies $S^{(m)}$. This appendix also records how the finite-$p$ values quoted there are evaluated.

The matrix elements have time averages in closed form. Each term of $V^{(00)}(m)$ and $W^{(00)}(m)$ in Eqs.~(\ref{eq:Vm}) and (\ref{eq:Wm}) is a product of two sinusoids in $m$, and reduces to sinusoids at the sum and difference frequencies $\Omega(\alpha \pm \gamma)$ and $2\Omega\gamma$ together with a constant part; the time average of $\cos(\omega m)$ over integer $m$ vanishes unless $\omega$ is a multiple of $2\pi$, which for these frequencies is precisely the phase-matching conditions (\ref{eq:cond_2}) through (\ref{eq:cond_4}). Away from those loci,
\begin{equation}
\lim_{p \to 0} V_\infty = \frac{1}{2},
\qquad
\lim_{p \to 0} W_\infty = -\frac{\alpha}{4\gamma^2},
\label{eq:VW_timeavg}
\end{equation}
so the time-averaged reduced state has eigenvalues $1/2 \pm \alpha/(4\gamma^2)$ and the averaged-state entropy approaches
\begin{equation}
\lim_{p \to 0} S(\bar\rho_A^{(\infty)}) = h\!\left(\frac{1}{2} + \frac{\alpha}{4\gamma^2}\right),
\label{eq:S_lin_limit}
\end{equation}
which at $\alpha = 1$ is $h(5/8) = 0.9544$ bits.

The trajectory-averaged entropy requires the single-realization concurrence, which follows from the reduced state alone: for a two-qubit pure state the eigenvalues of $\rho_A(m)$ are $(1 \pm \sqrt{1 - C_m^2})/2$, so their product gives $\det \rho_A(m) = C_m^2/4$ and
\begin{equation}
C_m = 2\sqrt{V(m)\left[1 - V(m)\right] - |W(m)|^2}.
\label{eq:Cm_det}
\end{equation}
The single-realization entropy $S^{(m)}$ of Eq.~(\ref{eq:Sm}) is therefore a bounded function of the two phases $(\Omega\alpha m, \Omega\gamma m)$, and for generic $\alpha$ this pair of phases, reduced modulo $2\pi$, fills its range uniformly~\cite{1916_Weyl_equidistribution}, so the time average of $S^{(m)}$ exists; a direct average over $2 \times 10^6$ steps, converged to the digits quoted, gives $0.4810$ bits at $\alpha = 1$.

Combining the two limits,
\begin{equation}
\lim_{p \to 0} \Delta S(\alpha, p) = h\!\left(\frac{1}{2} + \frac{\alpha}{4\gamma^2}\right) - \lim_{M\to\infty} \frac{1}{M} \sum_{m=0}^{M-1} S^{(m)},
\label{eq:DeltaS_limit}
\end{equation}
which is positive by the strict concavity of the entropy functional unless $\rho_A(m)$ is independent of $m$, and which at $\alpha = 1$ equals $0.4735$, the $0.47$ quoted in Sec.~\ref{subsec:DeltaS}. The finite-$p$ values quoted there follow from evaluating Eqs.~(\ref{eq:S_avgrho}) and (\ref{eq:Sbar_closedform}) on a grid of $p$ at $\alpha = 1$: the gap is maximal, $0.498$, at $p^* = 0.130$. The same construction applies to the Bell reset, for which Eq.~(\ref{eq:Winfbell}) gives $\lim_{p \to 0} W_\infty = -\alpha/(2\gamma^2)$, the averaged-state entropy approaches $h\bigl(\tfrac{1}{2} + \alpha/(2\gamma^2)\bigr)$, which is $h(3/4) = 0.8113$ bits at $\alpha = 1$, and the limit of the gap is $0.14$ bits.

\section{Concurrence at fixed age for the two reset states}
\label{app:trajectory_avg_entropy}

The single-realization concurrence $C_m$ of Eq.~(\ref{eq:Cm}) requires the amplitudes of the state at age $m$ in the computational basis, which this appendix gives for the product-state reset, together with the analogous construction for the Bell reset. Both feed the trajectory-averaged entropy of Eq.~(\ref{eq:Sbar_closedform}).

For the product-state reset $\ket{\psi_0} = \ket{00}$, a single realization at age $m$ is the pure state $\ket{\psi(m)} = U^m \ket{00}$. Expanding in the eigenbasis of $H$ and reading the overlaps from the decomposition (\ref{eq:overlap_00}) of $\ket{00}$ and the eigenstates (\ref{eq:eigenstates_plus_E3}) and (\ref{eq:eigenstates_plus}),
\begin{equation}
\ket{\psi(m)} = \sum_{k} e^{-i E_k m}\,\braket{E_k|00}\,\ket{E_k},
\label{eq:psi_m_eigenbasis}
\end{equation}
and projecting onto the computational basis, with $\cos\theta = \alpha/\gamma$ and $\sin\theta = 1/\gamma$ for the mixing angle of Sec.~\ref{sec:resonances}, gives the amplitudes $c_{ab}(m) = \braket{ab|\psi(m)}$ of Eq.~(\ref{eq:Cm}),
\begin{align}
c_{00}(m) &= \tfrac{1}{2}\bigl[\,e^{iJm} + \cos(\Omega\gamma m) + i\,(\alpha/\gamma)\sin(\Omega\gamma m)\,\bigr],
\notag \\
c_{11}(m) &= \tfrac{1}{2}\bigl[-e^{iJm} + \cos(\Omega\gamma m) + i\,(\alpha/\gamma)\sin(\Omega\gamma m)\,\bigr],
\notag \\
c_{01}(m) &= c_{10}(m) = -\frac{i}{2\gamma}\,\sin(\Omega\gamma m).
\label{eq:amplitudes_00}
\end{align}
The odd-sector component $\ket{\Phi^-}$ of $\ket{00}$ contributes the phase $e^{iJm}$, which enters $c_{00}$ and $c_{11}$ with opposite signs, and the even-sector component contributes the phase $\Omega\gamma m$; the equal amplitudes $c_{01} = c_{10}$ reflect the exchange symmetry of $H$. Because the reduced state of one qubit determines the pure-state concurrence through $\det \rho_A(m) = C_m^2/4$, the combination in Eq.~(\ref{eq:Cm}) equals the determinant form of Eq.~(\ref{eq:Cm_det}), written in terms of the reduced-state entries $V(m)$ and $W(m)$ of Eqs.~(\ref{eq:Vm}) and (\ref{eq:Wm}). Equation~(\ref{eq:Cm_det}) is the expression evaluated in Eq.~(\ref{eq:Sbar_closedform}).

For the Bell reset $\ket{\psi_0} = \ket{\Phi^+}$, the reduced state at age $m$ has $V^{(\Phi)}(m) = 1/2$ and $W^{(\Phi)}(m) = -\alpha \sin^2(\Omega\gamma m)/\gamma^2$ from Eq.~(\ref{eq:Wmbell}), so Eq.~(\ref{eq:Cm_det}) gives
\begin{equation}
C_m^{\mathrm{Bell}} = 2\sqrt{\tfrac{1}{4} - |W(m)|^2}
                    = \sqrt{1 - \frac{4\,\alpha^2 \sin^4(\Omega\gamma m)}{\gamma^4}}.
\label{eq:Cm_bell}
\end{equation}
Substituting (\ref{eq:Cm_bell}) into Eq.~(\ref{eq:Sm}) and then into Eq.~(\ref{eq:Sbar_closedform}) gives the Bell-reset trajectory-averaged entropy. At $\Omega\gamma = \pi$ the factor $\sin(\Omega\gamma m)$ vanishes for every integer $m$, so $C_m^{\mathrm{Bell}} = 1$ at all ages and every realization is the pure Bell state, consistent with the vanishing of the Jensen gap on the circular locus.

\section{First-order noise response of the two diagnostics}
\label{app:noise_perturbation}

The first-order correction to the noisy stationary state of Eq.~(\ref{eq:noisy_fixedpoint}), and the slopes $|\partial S/\partial\epsilon|$ and $|\partial C/\partial\epsilon|$ quoted in Sec.~\ref{subsec:structural_origin}, follow from writing the state as $\bar\rho_\infty^{(\epsilon)} = \bar\rho_\infty + \epsilon\,\bar\rho_1 + O(\epsilon^2)$ and the channel as $\mathcal{E}_\epsilon = \mathcal{I} + \epsilon\,\mathcal{D}' + O(\epsilon^2)$. In the Pauli basis $\mathcal{D}'$ is diagonal and multiplies a weight-$w$ string by $-w$, so a single application damps the weight-one coefficients at rate one and the weight-two coefficients at rate two, as stated in Sec.~\ref{subsec:structural_origin}. Substituting both expansions into Eq.~(\ref{eq:noisy_fixedpoint}) and matching orders, the zeroth order reproduces the noiseless fixed point (\ref{eq:fixedpoint}), and the first order is
\begin{equation}
\bar\rho_1 = (1-p)\,\mathcal{U}[\bar\rho_1] + (1-p)\,\mathcal{D}'\!\bigl[\mathcal{U}[\bar\rho_\infty]\bigr],
\qquad \mathcal{U}[X] \equiv U X U^\dagger ,
\label{eq:noise_firstorder}
\end{equation}
with the solution
\begin{equation}
\bar\rho_1 = (1-p)\,\bigl[\mathcal{I} - (1-p)\,\mathcal{U}\bigr]^{-1}
              \mathcal{D}'\!\bigl[\mathcal{U}[\bar\rho_\infty]\bigr].
\label{eq:noise_firstorder_soln}
\end{equation}
The inverse is well defined for every $p > 0$, and near the noiseless resonances of Sec.~\ref{sec:resonances} two further eigenvalues of the operator $\mathcal{I} - (1-p)\mathcal{U}$, beyond the value $p$ it takes on the populations at every parameter value, approach $p$, so $\bar\rho_1$ is large there. This is why the entropy slope of the product-state reset and the concurrence slope of the Bell reset evaluated below are largest on the circular locus (\ref{eq:cond_2}).

The slopes follow by the chain rule. Because $\bar\rho_1$ is traceless, the averaged-state entropy slope is
\begin{equation}
\frac{\partial S}{\partial\epsilon}\Big|_{\epsilon=0}
= -\operatorname{Tr}\!\bigl[\,\log_2 \bar\rho_A \;\operatorname{Tr}_{B}\bar\rho_1\,\bigr] ,
\label{eq:dSdeps}
\end{equation}
which depends on $\bar\rho_1$ only through its reduced state $\operatorname{Tr}_B\bar\rho_1$, that is, through the identity and the weight-one content that survives the partial trace. The concurrence slope depends on all sixteen Pauli components of $\bar\rho_1$, including the weight-two content, and is taken as $\epsilon \to 0^+$, since $C$ is not differentiable where a Wootters eigenvalue vanishes. The propagator $[\mathcal{I}-(1-p)\mathcal{U}]^{-1}$ mixes the Pauli-weight sectors, so the per-step grading of $\mathcal{D}'$ does not transfer to $\bar\rho_1$ unchanged; the suppression of the entropy response relative to the concurrence is a difference of rates established numerically from Eqs.~(\ref{eq:dSdeps}) and (\ref{eq:noise_firstorder_soln}).

Evaluating Eqs.~(\ref{eq:noise_firstorder_soln}) and (\ref{eq:dSdeps}) across the Bell-reset scan of Fig.~\ref{fig:S_Sbar_DeltaS}, $p \in \{0.05, 0.35, 0.65, 0.95\}$ and $\alpha \in [0, 4]$, and across the product-state reset at $p = 0.1$ for $0 < \alpha \leq 4$ (at $\alpha = 0$ the product-state reset gives a separable stationary state, so the concurrence and its slope both vanish) gives $|\partial S/\partial\epsilon| \leq 0.33\,|\partial C/\partial\epsilon|$, the ratio reaching its maximum for the Bell reset at $p = 0.05$ near $\alpha = 0.9$. At the operating point of Fig.~\ref{fig.1:circuit} ($\alpha = 3.2$, $p = 0.1$) the product-state reset has $|\partial S/\partial\epsilon| = 0.95$ bits and $|\partial C/\partial\epsilon| = 12.7$. The entropy slope is largest on the circular locus (\ref{eq:cond_2}): for the product-state reset at $p = 0.1$ it reaches $1.46$ bits at $\alpha = \sqrt{\pi^2 - 1}$. Away from the locus it is below $0.03$ bits for $\alpha \leq 0.1$ and near $\alpha = 2.25$, it reaches $1.05$ bits at $\alpha = 1.0$, and it rises to $1.46$ bits at $\alpha = 4$. For the Bell reset at the same locus $U\ket{\Phi^+} = -\ket{\Phi^+}$, the reduced state is maximally mixed and $\partial S/\partial\epsilon$ vanishes identically, while $|\partial C/\partial\epsilon| = 27$; this is the extremal case of the asymmetry.

\section{Phase matching for the chain}
\label{app:chain_phase_matching}

The resonance classification of Sec.~\ref{sec:resonances} is performed at two qubits, but its structure extends to the chain because the transverse-field Ising model is integrable. A Jordan-Wigner transformation maps Eq.~(\ref{eq:chain_hamiltonian}) to free fermions with single-particle energies $\epsilon_q(J, \Omega)$ \cite{1970_Pfeuty_ising}, and every many-body eigenenergy is a constant plus a sum $\sum_q \nu_q\,\epsilon_q$ over mode occupations $\nu_q \in \{0, 1\}$. The eigenenergy differences that enter the resonance condition $(E_i - E_j)T \in 2\pi\mathbb{Z}$ are therefore integer combinations $\sum_q m_q\,\epsilon_q(J, \Omega) \in 2\pi\mathbb{Z}$ of the single-particle energies, and the reset-state overlap $\bra{E_i}\rho_0\ket{E_j}$ selects which combinations contribute, generalizing the parity selection rule of Sec.~\ref{subsec:selection} to the fermion-parity and mode-occupation sectors on which $\ket{0}^{\otimes N}$ and the tiled Bell state have support.

The resonance loci are the conditions $\sum_q m_q\,\epsilon_q(J,\Omega)\,T \in 2\pi\mathbb{Z}$ on the single-particle energies; at $N = 2$ the two fermion modes $\epsilon_{1,2} = \Omega(\gamma \pm \alpha)$ give the four combinations $\epsilon_1$, $\epsilon_2$, $\epsilon_1 + \epsilon_2$, and $\epsilon_1 - \epsilon_2$ that reproduce Eqs.~(\ref{eq:cond_4}), (\ref{eq:cond_3}), (\ref{eq:cond_2}), and (\ref{eq:cond_1}) respectively. The mode wave numbers of the open chain, in turn, are spaced by order $1/N$, so the number of loci grows with $N$, and as $N \to \infty$ the single-particle band fills an interval and the discrete loci merge into continuous resonance bands in the $(\Omega, J)$ plane.

The half-chain Jensen gap of Sec.~\ref{sec:exact_chain} does not depend on this resonance structure. It equals the Holevo information of Eq.~(\ref{eq:holevo}), a property of the full family of reduced states rather than of any single resonant coherence, so the scaling reported there does not rest on the resonance structure. A momentum-resolved classification of the chain resonances and their imprint on the averaged-state entropy at chain scale is left to future work.

\bibliography{references}

@article{2025_Yin_resonances,
  title   = {Resonances of recurrence time of monitored quantum walks},
  author  = {Yin, Ruoyu and Wang, Qingyuan and Tornow, Sabine and Barkai, Eli},
  journal = {J. Chem. Phys.},
  volume  = {162},
  pages   = {244114},
  year    = {2025},
  doi     = {10.1063/5.0265944},
  url     = {https://doi.org/10.1063/5.0265944},
}

@article{2026_Liu_recurrence,
  title   = {Fractionally quantized recurrence detection times in monitored quantum many-body systems},
  author  = {Liu, Quancheng and Tornow, Sabine and Kessler, David A. and Barkai, Eli},
  journal = {Proc. Natl. Acad. Sci. U.S.A.},
  volume  = {123},
  number  = {22},
  pages   = {e2529694123},
  year    = {2026},
  doi     = {10.1073/pnas.2529694123},
  url     = {https://doi.org/10.1073/pnas.2529694123},
}

@article{2023_Singh_Bernoulli_reset,
  title = {Bernoulli trial under restarts: A comparative study of resetting transitions},
  author = {Singh, R. K. and Sandev, T. and Singh, Sadhana},
  journal = {Phys. Rev. E},
  volume = {108},
  issue = {5},
  pages = {L052106},
  numpages = {6},
  year = {2023},
  month = {Nov},
  publisher = {American Physical Society},
  doi = {10.1103/PhysRevE.108.L052106},
  url = {https://link.aps.org/doi/10.1103/PhysRevE.108.L052106}
}

@article{2026_Singh_splitting,
  title = {Transition in Splitting Probabilities of Quantum Walks},
  author = {Singh, Prashant and Kessler, David A. and Barkai, Eli},
  journal = {Phys. Rev. Lett.},
  volume = {137},
  issue = {9},
  pages = {090405},
  numpages = {9},
  year = {2026},
  month = {Aug},
  publisher = {American Physical Society},
  doi = {10.1103/gvzq-n573},
  url = {https://link.aps.org/doi/10.1103/gvzq-n573}
}

@misc{2025_Soldner_nonanalyticities,
      title={Nonanalyticities and ergodicity breaking in noninteracting many-body dynamics via stochastic resetting and global measurements}, 
      author={David Soldner and Igor Lesanovsky and Gabriele Perfetto},
      year={2025},
      eprint={2510.11450},
      archivePrefix={arXiv},
      primaryClass={cond-mat.stat-mech},
      url={https://arxiv.org/abs/2510.11450}, 
}

@article{2025_Kulkarni_bosons,
  author  = {Kulkarni, Manas and Majumdar, Satya N. and Sabhapandit, Sanjib},
  title   = {Dynamically emergent correlations in bosons via quantum resetting},
  journal = {J. Phys. A: Math. Theor.},
  volume  = {58},
  pages   = {105003},
  year    = {2025},
  doi     = {10.1088/1751-8121/adb6db},
  url     = {https://doi.org/10.1088/1751-8121/adb6db},
}

@misc{2026_Ma_noisy,
  author        = {Ma, Shuanger and Tornow, Sabine and Barkai, Eli},
  title         = {Resonances, recurrence times and steady states in monitored noisy qubit systems},
  year          = {2026},
  eprint        = {2603.18996},
  archivePrefix = {arXiv},
  primaryClass  = {cond-mat.stat-mech},
  url           = {https://arxiv.org/abs/2603.18996},
}

@misc{2026_Gotta_scars,
      title={Towers of quantum many-body scars under stochastic resetting}, 
      author={Lorenzo Gotta and Manas Kulkarni and Gabriele Perfetto},
      year={2026},
      eprint={2603.13165},
      archivePrefix={arXiv},
      primaryClass={cond-mat.stat-mech},
      url={https://arxiv.org/abs/2603.13165}, 
}

@article{2026_Jafari_defects,
  author  = {Jafari, R. and Johannesson, Henrik and Eggert, Sebastian},
  title   = {Topological and nontopological defects from quantum reset dynamics},
  journal = {Phys. Rev. B},
  volume  = {113},
  pages   = {224312},
  year    = {2026},
  doi     = {10.1103/PhysRevB.113.224312},
  url     = {https://doi.org/10.1103/PhysRevB.113.224312},
}

@article{2025_Yin_uncertainty,
  author  = {Yin, Ruoyu and Wang, Qingyuan and Tornow, Sabine and Barkai, Eli},
  title   = {Restart uncertainty relation for monitored quantum dynamics},
  journal = {Proc. Natl. Acad. Sci. U.S.A.},
  volume  = {122},
  number  = {1},
  pages   = {e2402912121},
  year    = {2025},
  doi     = {10.1073/pnas.2402912121},
  url     = {https://doi.org/10.1073/pnas.2402912121},
}

@article{2026_Hamazaki_review,
  author  = {Hamazaki, Ryusuke and Mochizuki, Ken and Oshima, Hisanori and
             Fuji, Yohei},
  title   = {An introduction to monitored quantum systems and quantum
             trajectories: spectrum, typicality, and phases},
  journal = {Prog. Theor. Exp. Phys.},
  volume  = {2026},
  number  = {5},
  pages   = {052A01},
  year    = {2026},
  doi     = {10.1093/ptep/ptag055},
  url     = {https://doi.org/10.1093/ptep/ptag055},
}

@article{2025_Kamakari_XEB,
  author  = {Kamakari, Hirsh and Sun, Jiace and Li, Yaodong and Thio,
             Jonathan J. and Gujarati, Tanvi P. and Fisher, Matthew P. A. and
             Motta, Mario and Minnich, Austin J.},
  title   = {Experimental demonstration of scalable cross-entropy benchmarking
             to detect measurement-induced phase transitions on a
             superconducting quantum processor},
  journal = {Phys. Rev. Lett.},
  volume  = {134},
  pages   = {120401},
  year    = {2025},
  doi     = {10.1103/PhysRevLett.134.120401},
  url     = {https://doi.org/10.1103/PhysRevLett.134.120401},
}

@article{2025_Qian_protect,
  author  = {Qian, Dongheng and Wang, Jing},
  title   = {Protect measurement-induced phase transition from noise},
  journal = {Phys. Rev. Lett.},
  volume  = {134},
  pages   = {020403},
  year    = {2025},
  doi     = {10.1103/PhysRevLett.134.020403},
  url     = {https://doi.org/10.1103/PhysRevLett.134.020403},
}

@book{1949_Hardy_divergent,
  author    = {Hardy, Godfrey H.},
  title     = {Divergent Series},
  publisher = {Oxford University Press},
  address   = {Oxford},
  year      = {1949}
}

@article{1916_Weyl_equidistribution,
author = {Weyl, Hermann},
date = {1916/09/01},
doi = {10.1007/BF01475864},
id = {Weyl1916},
isbn = {1432-1807},
journal = {Math. Ann.},
number = {3},
pages = {313--352},
title = {{\"U}ber die {Gleichverteilung} von {Zahlen} mod. {Eins}},
url = {https://doi.org/10.1007/BF01475864},
volume = {77},
year = {1916}}

@article{2021_Corcoles_dynamic,
  title = {Exploiting Dynamic Quantum Circuits in a Quantum Algorithm with Superconducting Qubits},
  author = {C{\'o}rcoles, A. D. and Takita, Maika and Inoue, Ken and Lekuch, Scott and Minev, Zlatko K. and Chow, Jerry M. and Gambetta, Jay M.},
  journal = {Phys. Rev. Lett.},
  volume = {127},
  issue = {10},
  pages = {100501},
  numpages = {6},
  year = {2021},
  month = {Aug},
  publisher = {American Physical Society},
  doi = {10.1103/PhysRevLett.127.100501},
  url = {https://link.aps.org/doi/10.1103/PhysRevLett.127.100501}
}

@article{2009_Briet_jensen,
  author  = {Bri{\"e}t, Jop and Harremo{\"e}s, Peter},
  title   = {Properties of classical and quantum {Jensen-Shannon} divergence},
  journal = {Phys. Rev. A},
  volume  = {79},
  pages   = {052311},
  year    = {2009},
  doi     = {10.1103/PhysRevA.79.052311},
  url     = {https://doi.org/10.1103/PhysRevA.79.052311},
  note    = {arXiv:0806.4472},
}

@article{2009_Verstraete_dissipation,
  author  = {Verstraete, Frank and Wolf, Michael M. and Cirac, J. Ignacio},
  title   = {Quantum computation and quantum-state engineering driven by dissipation},
  journal = {Nat. Phys.},
  volume  = {5},
  pages   = {633--636},
  year    = {2009},
  doi     = {10.1038/nphys1342},
  url     = {https://doi.org/10.1038/nphys1342},
  note    = {arXiv:0803.1447},
}

@book{2010_Nielsen_qcqi,
  title     = {{Quantum Computation and Quantum Information}: 10th Anniversary Edition},
  author    = {Nielsen, Michael A. and Chuang, Isaac L.},
  publisher = {Cambridge University Press},
  year      = {2010},
  doi       = {10.1017/CBO9780511976667},
  url       = {https://doi.org/10.1017/CBO9780511976667},
}

@article{1978_Wehrl_entropy,
  author  = {Wehrl, Alfred},
  title   = {General properties of entropy},
  journal = {Rev. Mod. Phys.},
  volume  = {50},
  number  = {2},
  pages   = {221--260},
  year    = {1978},
  doi     = {10.1103/RevModPhys.50.221},
  url     = {https://doi.org/10.1103/RevModPhys.50.221},
}

@misc{2026_IBM_platform,
  author       = {{IBM Quantum}},
  title        = {{IBM} Quantum Platform},
  howpublished = {\url{https://quantum.cloud.ibm.com/}},
  note         = {Accessed: 4 July 2026},
  year         = {2026},
}

@misc{2026_Murauer_NESS,
      title={Nonequilibrium steady states induced by stochastic mid-circuit measurements and resets on a quantum computer}, 
      author={Jakob Murauer and Sabine Tornow and Gabriele Perfetto},
      year={2026},
      eprint={2606.19027},
      archivePrefix={arXiv},
      primaryClass={quant-ph},
      url={https://arxiv.org/abs/2606.19027}, 
}

@article{2026_Ghosh_concurrence,
  title = {Generating pairwise entanglement in periodically driven quantum spin chains with stochastic resetting},
  author = {Ghosh, Sinchan and Kulkarni, Manas and Sengupta, K. and Majumdar, Satya N.},
  journal = {Phys. Rev. B},
  volume = {114},
  issue = {14},
  pages = {144406},
  numpages = {7},
  year = {2026},
  month = {Sep},
  publisher = {American Physical Society},
  doi = {10.1103/k58k-tpy7},
  url = {https://link.aps.org/doi/10.1103/k58k-tpy7}
}

@misc{2026_Yokoyama_resetcircuits,
      title={Continuous Reset-Induced Phase Transition in Measurement-Free Random Quantum Circuits}, 
      author={Hinata Yokoyama and Kengo Anzai and Dina Syverud-Lindland and Yoshihito Kuno and Hiroaki Matsueda},
      year={2026},
      eprint={2604.25640},
      archivePrefix={arXiv},
      primaryClass={quant-ph},
      url={https://arxiv.org/abs/2604.25640}, 
}

@misc{2026_Das_lackadaisical,
      title={Restart and first detection in a lackadaisical quantum walk with flat-band localization}, 
      author={Debraj Das},
      year={2026},
      eprint={2609.08973},
      archivePrefix={arXiv},
      primaryClass={quant-ph},
      url={https://arxiv.org/abs/2609.08973}, 
}

@article{2026_Jafari_resetenv,
  author  = {Jafari, R. and Asadian, Ali and Biderang, Mehdi and Akbari, Alireza},
  title   = {Quantum correlation dynamics subjected to quantum reset-driven environment},
  journal = {J. Phys. A: Math. Theor.},
  volume  = {59},
  pages   = {355301},
  year    = {2026},
  doi     = {10.1088/1751-8121/ae9b50},
  url     = {https://doi.org/10.1088/1751-8121/ae9b50},
}

@misc{2022_Ragonneau_dfo,
      title={Model-Based Derivative-Free Optimization Methods and Software}, 
      author={Tom M. Ragonneau},
      year={2023},
      eprint={2210.12018},
      archivePrefix={arXiv},
      primaryClass={math.OC},
      url={https://arxiv.org/abs/2210.12018}, 
}

@misc{2025_Ragonneau_cobyqa,
  author  = {Ragonneau, Tom M. and Zhang, Zaikun},
  title   = {{COBYQA} version 1.1.3},
  year    = {2025},
  url     = {https://www.cobyqa.com},
}

@article{1973_Holevo_bound,
  author  = {Holevo, A. S.},
  title   = {Bounds for the quantity of information transmitted by a
             quantum communication channel},
  journal = {Probl. Inf. Transm.},
  volume  = {9},
  number  = {3},
  pages   = {177--183},
  year    = {1973},
  note    = {Translated from Probl. Peredachi Inf. 9(3), 3--11 (1973)},
  url     = {https://www.mathnet.ru/eng/ppi903},
}

@article{1997_Hill_concurrence,
  title   = {Entanglement of a pair of quantum bits},
  author  = {Hill, Sam A. and Wootters, William K.},
  journal = {Phys. Rev. Lett.},
  volume  = {78},
  pages   = {5022--5025},
  year    = {1997},
  doi     = {10.1103/PhysRevLett.78.5022},
  url     = {https://doi.org/10.1103/PhysRevLett.78.5022},
}

@article{1998_Wootters_concurrence,
  title   = {Entanglement of formation of an arbitrary state of two qubits},
  author  = {Wootters, William K.},
  journal = {Phys. Rev. Lett.},
  volume  = {80},
  pages   = {2245},
  year    = {1998},
  doi     = {10.1103/PhysRevLett.80.2245},
  url     = {https://doi.org/10.1103/PhysRevLett.80.2245},
}

@article{2009_Horodecki_review,
  title   = {Quantum entanglement},
  author  = {Horodecki, Ryszard and Horodecki, Pawe{\l} and Horodecki, Micha{\l} and Horodecki, Karol},
  journal = {Rev. Mod. Phys.},
  volume  = {81},
  pages   = {865--942},
  year    = {2009},
  doi     = {10.1103/RevModPhys.81.865},
  url     = {https://doi.org/10.1103/RevModPhys.81.865},
}

@article{2011_Evans_diffusion,
  title   = {Diffusion with stochastic resetting},
  author  = {Evans, Martin R. and Majumdar, Satya N.},
  journal = {Phys. Rev. Lett.},
  volume  = {106},
  pages   = {160601},
  year    = {2011},
  doi     = {10.1103/PhysRevLett.106.160601},
  url     = {https://doi.org/10.1103/PhysRevLett.106.160601},
}

@article{2013_Grunbaum_recurrence,
  author  = {Gr{\"u}nbaum, F. A. and Vel{\'a}zquez, L. and Werner, A. H. and Werner, R. F.},
  title   = {Recurrence for discrete time unitary evolutions},
  journal = {Commun. Math. Phys.},
  volume  = {320},
  number  = {2},
  pages   = {543--569},
  year    = {2013},
  doi     = {10.1007/s00220-012-1645-2},
  url     = {https://doi.org/10.1007/s00220-012-1645-2},
}

@article{2017_Friedman_walks,
  title   = {Quantum walks: the first detected passage time problem},
  author  = {Friedman, H. and Kessler, D. A. and Barkai, E.},
  journal = {Phys. Rev. E},
  volume  = {95},
  pages   = {032141},
  year    = {2017},
  doi     = {10.1103/PhysRevE.95.032141},
  url     = {https://doi.org/10.1103/PhysRevE.95.032141},
}

@article{2017_Pal_restart,
  title   = {First passage under restart},
  author  = {Pal, Arnab and Reuveni, Shlomi},
  journal = {Phys. Rev. Lett.},
  volume  = {118},
  pages   = {030603},
  year    = {2017},
  doi     = {10.1103/PhysRevLett.118.030603},
  url     = {https://doi.org/10.1103/PhysRevLett.118.030603},
}

@article{2018_Mukherjee_reset,
  title   = {Quantum dynamics with stochastic reset},
  author  = {Mukherjee, B. and Sengupta, K. and Majumdar, Satya N.},
  journal = {Phys. Rev. B},
  volume  = {98},
  pages   = {104309},
  year    = {2018},
  doi     = {10.1103/PhysRevB.98.104309},
  url     = {https://doi.org/10.1103/PhysRevB.98.104309},
}

@article{2018_Rose_spectral,
  title   = {Spectral properties of simple classical and quantum reset processes},
  author  = {Rose, Dominic C. and Touchette, Hugo and Lesanovsky, Igor and Garrahan, Juan P.},
  journal = {Phys. Rev. E},
  volume  = {98},
  pages   = {022129},
  year    = {2018},
  doi     = {10.1103/PhysRevE.98.022129},
  url     = {https://doi.org/10.1103/PhysRevE.98.022129},
}

@article{2019_Skinner_MIPT,
  title   = {Measurement-induced phase transitions in the dynamics of entanglement},
  author  = {Skinner, Brian and Ruhman, Jonathan and Nahum, Adam},
  journal = {Phys. Rev. X},
  volume  = {9},
  pages   = {031009},
  year    = {2019},
  doi     = {10.1103/PhysRevX.9.031009},
  url     = {https://doi.org/10.1103/PhysRevX.9.031009},
}

@article{2019_Li_hybrid,
  title   = {Measurement-driven entanglement transition in hybrid quantum circuits},
  author  = {Li, Yaodong and Chen, Xiao and Fisher, Matthew P. A.},
  journal = {Phys. Rev. B},
  volume  = {100},
  pages   = {134306},
  year    = {2019},
  doi     = {10.1103/PhysRevB.100.134306},
  url     = {https://doi.org/10.1103/PhysRevB.100.134306},
}

@article{2020_Gullans_purification,
  title   = {Dynamical purification phase transition induced by quantum measurements},
  author  = {Gullans, Michael J. and Huse, David A.},
  journal = {Phys. Rev. X},
  volume  = {10},
  pages   = {041020},
  year    = {2020},
  doi     = {10.1103/PhysRevX.10.041020},
  url     = {https://doi.org/10.1103/PhysRevX.10.041020},
}

@article{2020_Evans_resetting,
  author  = {Evans, Martin R. and Majumdar, Satya N. and Schehr, Gr{\'e}gory},
  title   = {Stochastic resetting and applications},
  journal = {J. Phys. A: Math. Theor.},
  volume  = {53},
  number  = {19},
  pages   = {193001},
  year    = {2020},
  doi     = {10.1088/1751-8121/ab7cfe},
  url     = {https://doi.org/10.1088/1751-8121/ab7cfe},
}

@article{2021_Perfetto_resetting,
  title   = {Designing nonequilibrium states of quantum matter through stochastic resetting},
  author  = {Perfetto, Gabriele and Carollo, Federico and Magoni, Matteo and Lesanovsky, Igor},
  journal = {Phys. Rev. B},
  volume  = {104},
  pages   = {L180302},
  year    = {2021},
  doi     = {10.1103/PhysRevB.104.L180302},
  url     = {https://doi.org/10.1103/PhysRevB.104.L180302},
}

@article{2022_Magoni_correlations,
  title   = {Emergent quantum correlations and collective behavior in noninteracting quantum systems subject to stochastic resetting},
  author  = {Magoni, Matteo and Carollo, Federico and Perfetto, Gabriele and Lesanovsky, Igor},
  journal = {Phys. Rev. A},
  volume  = {106},
  pages   = {052210},
  year    = {2022},
  doi     = {10.1103/PhysRevA.106.052210},
  url     = {https://doi.org/10.1103/PhysRevA.106.052210},
}

@article{2022_Turkeshi_resetting,
  title   = {Entanglement transitions from stochastic resetting of non-{H}ermitian quasiparticles},
  author  = {Turkeshi, Xhek and Dalmonte, Marcello and Fazio, Rosario and Schir{\`o}, Marco},
  journal = {Phys. Rev. B},
  volume  = {105},
  pages   = {L241114},
  year    = {2022},
  doi     = {10.1103/PhysRevB.105.L241114},
  url     = {https://doi.org/10.1103/PhysRevB.105.L241114},
}

@article{2023_Tornow_walks,
  title   = {Measurement-induced quantum walks on an {IBM} quantum computer},
  author  = {Tornow, Sabine and Ziegler, Klaus},
  journal = {Phys. Rev. Res.},
  volume  = {5},
  pages   = {033089},
  year    = {2023},
  doi     = {10.1103/PhysRevResearch.5.033089},
  url     = {https://doi.org/10.1103/PhysRevResearch.5.033089},
}

@article{2023_Koh_MIPT,
  title   = {Measurement-induced entanglement phase transition on a superconducting quantum processor with mid-circuit readout},
  author  = {Koh, Jin Ming and Sun, Shi-Ning and Motta, Mario and Minnich, Austin J.},
  journal = {Nat. Phys.},
  volume  = {19},
  pages   = {1314--1319},
  year    = {2023},
  doi     = {10.1038/s41567-023-02076-6},
  url     = {https://doi.org/10.1038/s41567-023-02076-6},
}

@article{2023_Kulkarni_entanglement,
  title   = {Generating entanglement by quantum resetting},
  author  = {Kulkarni, Manas and Majumdar, Satya N.},
  journal = {Phys. Rev. A},
  volume  = {108},
  pages   = {062210},
  year    = {2023},
  doi     = {10.1103/PhysRevA.108.062210},
  url     = {https://doi.org/10.1103/PhysRevA.108.062210},
}

@article{2023_Kulkarni_detection,
	author = {Kulkarni, Manas and Majumdar, Satya N},
	doi = {10.1088/1751-8121/acf103},
	journal = {J. Phys. A: Math. Theor.},
	month = {sep},
	number = {38},
	pages = {385003},
	publisher = {IOP Publishing},
	title = {First detection probability in quantum resetting via random projective measurements},
	url = {https://doi.org/10.1088/1751-8121/acf103},
	volume = {56},
	year = {2023}}

@article{2024_AlcaldePuente_ancilla,
  title         = {Quantum state preparation via engineered ancilla resetting},
  author        = {Alcalde Puente, Daniel and Motzoi, Felix and Calarco, Tommaso and Morigi, Giovanna and Rizzi, Matteo},
  journal       = {Quantum},
  volume        = {8},
  pages         = {1299},
  year          = {2024},
  doi           = {10.22331/q-2024-03-27-1299},
  url           = {https://doi.org/10.22331/q-2024-03-27-1299},
  eprint        = {2305.08641},
  archivePrefix = {arXiv},
  primaryClass  = {quant-ph},
}

@misc{2026_King_search,
  author        = {King, Emma C. and Roy, Sayan and Mattiotti, Francesco and Kiefer-Emmanouilidis, Maximilian and Bl{\"a}ser, Markus and Morigi, Giovanna},
  title         = {Time complexity of a monitored quantum search with resetting},
  year          = {2026},
  eprint        = {2601.20560},
  archivePrefix = {arXiv},
  primaryClass  = {quant-ph},
  url           = {https://arxiv.org/abs/2601.20560},
}

@article{2023_Yin_restart,
  title   = {Restart expedites quantum walk hitting times},
  author  = {Yin, R. and Barkai, E.},
  journal = {Phys. Rev. Lett.},
  volume  = {130},
  pages   = {050802},
  year    = {2023},
  doi     = {10.1103/PhysRevLett.130.050802},
  url     = {https://doi.org/10.1103/PhysRevLett.130.050802},
}

@article{2025_Wald_discrete,
  title         = {Stochastic resetting in discrete-time quantum dynamics: steady states and correlations in few-qubit systems},
  author        = {Wald, Sascha and Yao, Louie Hong and Platini, Thierry and Hooley, Chris and Carollo, Federico},
  journal       = {Quantum},
  volume        = {9},
  pages         = {1742},
  year          = {2025},
  doi           = {10.22331/q-2025-05-13-1742},
  url           = {https://doi.org/10.22331/q-2025-05-13-1742},
}

@article{2025_Koh_readout,
  title = {Readout Error Mitigation for Mid-Circuit Measurements and Feedforward},
  author = {Koh, Jin Ming and Koh, Dax Enshan and Thompson, Jayne},
  journal = {PRX Quantum},
  volume = {7},
  issue = {1},
  pages = {010317},
  numpages = {40},
  year = {2026},
  month = {Jan},
  publisher = {American Physical Society},
  doi = {10.1103/cj89-4h5t},
  url = {https://link.aps.org/doi/10.1103/cj89-4h5t}
}

@misc{2026_Carollo_nonMarkovianity,
      title={Stochastic resetting induces quantum non-{M}arkovianity}, 
      author={Federico Carollo and Sascha Wald},
      year={2026},
      eprint={2601.13367},
      archivePrefix={arXiv},
      primaryClass={quant-ph},
      url={https://arxiv.org/abs/2601.13367}, 
}

@misc{2026_SinghRoy_code,
  author  = {Singh Roy, Monalisa},
  title   = {{QuantumResets.jl}: nonequilibrium steady states of interacting quantum spin chains under stochastic resetting and their entanglement diagnostics},
  year    = {2026},
  version = {1.0.0},
  url     = {https://github.com/monalisasroy/QuantumResets.jl}
}

@article{2001_Godreche_occupation,
  author  = {Godr\`eche, C. and Luck, J. M.},
  title   = {Statistics of the occupation time of renewal processes},
  journal = {J. Stat. Phys.},
  volume  = {104},
  number  = {3--4},
  pages   = {489--524},
  year    = {2001},
  doi     = {10.1023/A:1010364003250},
  url     = {https://doi.org/10.1023/A:1010364003250},
}

@book{1971_Feller_probability,
  author    = {Feller, William},
  title     = {An Introduction to Probability Theory and Its Applications, Vol. II},
  edition   = {2nd},
  publisher = {Wiley},
  address   = {New York},
  year      = {1971},
  note      = {Renewal theory and recurrence times, Chap. XI},
}

@article{2023_Barkai_ergodic,
  author  = {Barkai, E. and Flaquer-Galm\'es, R. and M\'endez, V.},
  title   = {Ergodic properties of {B}rownian motion under stochastic resetting},
  journal = {Phys. Rev. E},
  volume  = {108},
  number  = {6},
  pages   = {064102},
  year    = {2023},
  doi     = {10.1103/PhysRevE.108.064102},
  url     = {https://doi.org/10.1103/PhysRevE.108.064102},
}

@article{2023_Afek_anomalous,
  title = {Colloquium: Anomalous statistics of laser-cooled atoms in dissipative optical lattices},
  author = {Afek, Gadi and Davidson, Nir and Kessler, David A. and Barkai, Eli},
  journal = {Rev. Mod. Phys.},
  volume = {95},
  issue = {3},
  pages = {031003},
  numpages = {22},
  year = {2023},
  month = {Sep},
  publisher = {American Physical Society},
  doi = {10.1103/RevModPhys.95.031003},
  url = {https://link.aps.org/doi/10.1103/RevModPhys.95.031003}
}

@article{2020_Turkeshi_criticality,
  author  = {Turkeshi, X. and Fazio, R. and Dalmonte, M.},
  title   = {Measurement-induced criticality in ($2+1$)-dimensional hybrid quantum circuits},
  journal = {Phys. Rev. B},
  volume  = {102},
  number  = {1},
  pages   = {014315},
  year    = {2020},
  doi     = {10.1103/PhysRevB.102.014315},
  url     = {https://doi.org/10.1103/PhysRevB.102.014315},
}

@article{2021_Turkeshi_clicks,
  author  = {Turkeshi, X. and Biella, A. and Fazio, R. and Dalmonte, M. and Schir\'o, M.},
  title   = {Measurement-induced entanglement transitions in the quantum {I}sing chain: From infinite to zero clicks},
  journal = {Phys. Rev. B},
  volume  = {103},
  number  = {22},
  pages   = {224210},
  year    = {2021},
  doi     = {10.1103/PhysRevB.103.224210},
  url     = {https://doi.org/10.1103/PhysRevB.103.224210},
}

@misc{2024_Wolfram_mathematica,
  author       = {{Wolfram Research, Inc.}},
  title        = {Mathematica, Version 14.x},
  howpublished = {Champaign, IL},
  year         = {2024},
  url          = {https://www.wolfram.com/mathematica/},
}

@article{1970_Pfeuty_ising,
  author  = {Pfeuty, P.},
  title   = {The one-dimensional {I}sing model with a transverse field},
  journal = {Ann. Phys.},
  volume  = {57},
  number  = {1},
  pages   = {79--90},
  year    = {1970},
  doi     = {10.1016/0003-4916(70)90270-8},
  url     = {https://doi.org/10.1016/0003-4916(70)90270-8},
}
\end{document}